\documentclass[10pt,journal,compsoc]{IEEEtran}
\ifCLASSOPTIONcompsoc
  \usepackage[nocompress]{cite}
\else
  \usepackage{cite}
\fi
\usepackage{tikz}
 
\ifCLASSINFOpdf
\else
\fi
\usepackage{fancyhdr}
\usepackage{booktabs} 
\usepackage{amssymb}
\usepackage{multirow}
\usepackage{color,graphicx}
\usepackage{amsmath}
\usepackage{amssymb}
\usepackage{float}
\usepackage{color}
\usepackage[ruled]{algorithm2e}
\usepackage{enumerate}
\usepackage{syntonly}
\usepackage{multicol}
\usepackage{array}
\usepackage[tight,footnotesize]{subfigure}
\usepackage{url}
\usepackage{enumitem}

\usepackage{comment}
\begin{document}

\title{ScanSAT: Unlocking Static and Dynamic\\Scan Obfuscation}

\author{Lilas~Alrahis,~\IEEEmembership{Student~Member,~IEEE,}
        Muhammad~Yasin,~\IEEEmembership{Member,~IEEE,}
        Nimisha~Limaye,~\IEEEmembership{Student~Member,~IEEE,}
        Hani~Saleh,~\IEEEmembership{Senior~Member,~IEEE,}
        Baker~Mohammad,~\IEEEmembership{Senior~Member,~IEEE,}
        Mahmoud~Al-Qutayri,~\IEEEmembership{Senior~Member,~IEEE,}
        and~Ozgur~Sinanoglu,~\IEEEmembership{Senior~Member,~IEEE
}

\thanks{A preliminary version of this work was presented at ASPDAC 2019 Conference~\cite{aspdac19}.}
\IEEEcompsocitemizethanks{\IEEEcompsocthanksitem L. Alrahis, H. Saleh, B. Mohammad and M. Al-Qutayri are with the Department of Electrical and Computer Engineering, Khalifa University, Abu Dhabi 127788, UAE.\protect\\
E-mail: lilas.alrahis@ku.ac.ae
\IEEEcompsocthanksitem N. Limaye is with the Department of Electrical and Computer Engineering,
Tandon School of Engineering, New York University, New York City, NY 11201, USA.
E-mail: nimisha.limaye@nyu.edu
\IEEEcompsocthanksitem M. Yasin is with the Department of Electrical and Computer Engineering at Texas A\&M University, College Station, TX 77843, USA.
\IEEEcompsocthanksitem O. Sinanoglu is with the Engineering Division, New York University Abu
Dhabi, Abu Dhabi 129188, UAE.
}%
\thanks{Manuscript received March 14, 2019; revised August 29, 2019.}}


\IEEEtitleabstractindextext{%
\begin{abstract}
While financially advantageous, outsourcing key steps, such as testing, to potentially untrusted Outsourced Assembly and Test (OSAT) companies may pose a risk of compromising on-chip assets. Obfuscation of scan chains is a technique that hides the actual scan data from the untrusted testers; logic inserted between the scan cells, driven by a secret key, hides the transformation functions that map the scan-in stimulus (scan-out response) and the delivered scan pattern (captured response). While static scan obfuscation utilizes the same secret key, and thus, the same secret transformation functions throughout the lifetime of the chip, dynamic scan obfuscation updates the key periodically. In this paper, we propose ScanSAT: an attack that transforms a scan obfuscated circuit to its logic-locked version and applies the Boolean satisfiability (SAT) based attack, thereby extracting the secret key. We implement our attack, apply on representative scan obfuscation techniques, and show that ScanSAT can break both static and dynamic scan obfuscation schemes with 100\% success rate. Moreover, ScanSAT is effective even for large key sizes and in the presence of scan compression.
\end{abstract}
\begin{IEEEkeywords}
Obfuscated~Scan~Chains, Scan~Obfuscation, Scan~Locking, SAT~Attack, Logic~Locking.
\end{IEEEkeywords}}
\maketitle
\renewcommand{\headrulewidth}{0.0pt}
\thispagestyle{fancy}
\pagestyle{fancy}
\cfoot{
\copyright~2019 IEEE.
This is the author's version of the work. It is posted here for your personal use.
	Not for redistribution.\\
	The definitive Version of Record is published in IEEE Transactions on Emerging Topics in Computing (TETC), 2019.
}
\lhead{IEEE TRANSACTIONS ON EMERGING TOPICS IN COMPUTING,~Vol.~, No.~,}%
\IEEEdisplaynontitleabstractindextext
\IEEEpeerreviewmaketitle
\section{Introduction}
\label{sec:introduction}
\vspace{-0.07cm}
\IEEEPARstart{M}{ore} and more design houses are going fabless due to the ever increasing cost of Integrated Circuit (IC) manufacturing. Even those who hold onto their fabrication facilities are now outsourcing key steps such as testing to OSAT facilities~\cite{rostami2014primer_1,osat_timing}. 
Outsourcing the fabrication and testing processes to potentially untrusted parties raises concerns regarding IC piracy, reverse engineering, overproduction, Intellectual Property (IP) rights violation, and hardware Trojan insertion~\cite{rostami2014primer_1}. Among the Design-for-Trust (DfTr) solutions developed to prevent such hardware security threats, logic locking is a holistic solution for mitigating IC piracy, Trojan insertion, and overproduction, as it provides protection throughout the IC supply chain.

\subsection{Logic Obfuscation/Locking}
Logic locking hides the functionality of the design via the insertion of additional logic elements (key gates).\footnote{Obfuscation, locking, and encryption have been used interchangeably in the literature. In this paper, we use the former two terms interchangeably.} The purpose of adding key gates is to lock the circuit during the untrusted phases of the design and manufacturing process. These key gates are driven by key-bits (key inputs) that are stored in a tamper-proof memory on the chip. A valid key restores the correct functionality of the design, unlocking it. Combinational logic locking inserts combinational key gates such as XOR/XNORs~\cite{epic,yasin_TCAD_2016}, or multiplexers (MUXes)~\cite{JV-Tcomp-2013,karmakar2018encrypt} to lock a design. The true functionality of the locked netlist/chip is dictated by the secret logic locking key, and is thus protected with this key. This way, not only the design IP is protected but also unauthorized use of the chips is prevented. The foundry, OSAT, and the end-users are all untrusted. 

\subsection{Scan Obfuscation/Locking}

Scan-based Design-for-Test (DfT) structures are used in manufacturing testing due to direct access to deeply embedded logic, enhancing test quality. However, this type of testing structures opens a side channel that can be utilized to retrieve secret information stored on-chip from cryptosystems~\cite{yang2006secure,nara2010scan,da2012new}. These attacks, referred to as {\em scan-based attacks}, exploit the scan-based test structure in order to steal security critical assets.

In order to provide protection against scan-based attacks as well as an untrusted OSAT company during the testing and chip-configuration phases, a special instance of logic locking, namely scan locking~\cite{karmakar2018encrypt,wang2017secure}, obfuscates the scan chain(s) by inserting key-driven logic in between the Scan Flip-Flops (SFFs). This way, the untrusted tester applies a scan-in stimulus (or configuration data) that is different than the pattern delivered into the scan chains; similarly, the tester observes scan-out responses that are different than the captured responses. Both transformations--the one that maps the inserted scan stimulus (or configuration data) to the pattern delivered into the scan chains, and the one that maps the captured response to the scan-out response--depend on the secret key, and are thus secret. The promise of scan locking is that so long as the key is kept secret, the actual test/configuration data cannot be inferred from the transformed (obfuscated) test/configuration data that the untrusted party has access to. Scan locking also prevents reverse engineering attempts that rely on the scan side channel~\cite{azriel2017using}. The OSAT and the end-users are untrusted, while the foundry may or may not be trusted depending on the threat and business models.

\color{black}
We would like to note that scan obfuscation refers to any technique that uses a set of keys to obfuscate the scan-in and scan-out data. Static scan obfuscation utilizes the same secret key, and thus, the same secret transformation functions throughout the lifetime of the chip~\cite{karmakar2018encrypt,hely2004scan}; the former technique inserts key-gates on the scan path, while the latter one reorders scan chain fragments based on a secret key. Dynamic scan obfuscation, on the other hand, updates the key periodically; the transformation function may be different for different test/configuration patterns~\cite{wang2017secure}. VIm-Scan~\cite{paul2007vim} and SSTKR~\cite{sstkr} can be considered dynamic obfuscation techniques that embed keys in test patterns, while \cite{wang2017secure} can be viewed as a generalized version of VIm-Scan and SSTKR techniques with the added flexibility of test data manipulations (pattern deletion/reordering). Furthermore, VIm-Scan and SSTKR are vulnerable to attacks that operate on test data to identify the key bits from the test patterns, and solve for the seed subsequently~\cite{wang2017secure}; this is not due to the weakness of these techniques, but rather because of the strong threat model that scan locking (and we) follows where the structural details about the design are available to the attacker. 

\begin{figure*}[!t]
	\centering
	\includegraphics[width=0.8\textwidth]{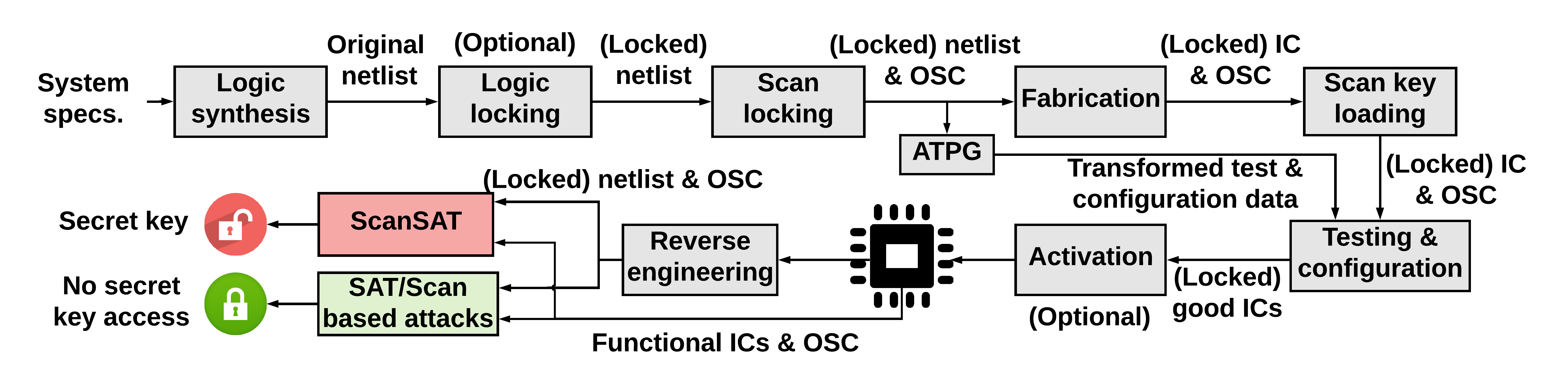}
	\caption{Design flow with logic and scan locking, and the proposed ScanSAT attack. OSC stands for Obfuscated Scan Chain. ATPG refers to Automatic Test Pattern Generation.}
	\label{fig:threat_model}
\end{figure*}

\vspace{-0.15in}
\subsection{Design Flow}
The design flow for an IC is demonstrated in Fig.~\ref{fig:threat_model}; it shows both logic locking and scan obfuscation. As the main focus of this work is an attack on scan obfuscation, we show logic locking in the design flow as an optional step; as shown later, scan obfuscation in conjunction with a basic logic locking scheme is still vulnerable to our attack. As the design flow shows, the netlist changes for both logic and scan locking can be done post-synthesis, while logic locking can also be integrated at RT Level~\cite{epic_journal,harpoon}. Scan locking enables a protocol where the designer loads the secret key post-manufacturing on some secure memory. The designer also generates the transformed (obfuscated) test data based on the secret key and the original test/configuration patterns. The transformed test data is provided to the OSAT company, who performs the testing without knowing the actual test patterns~\cite{yasin_DATE_2016,karmakar2018encrypt}. Similarly, the designer provides transformed configuration vectors that need to be delivered through the scan chains; the OSAT company applies the configuration vectors to customize each IC without being able to infer the actual content (security-critical bit streams, chip ID, etc.). Post-testing, the logic-locked chips that have been identified to be defect-free are activated by loading the secret logic locking key on the tamper-proof memory.\footnote{A chip need not be functional to be tested for structural defects. Test generation and application can thus be performed based on dummy logic locking keys. The flow therefore shows post-test activation of chips, as this is the most secure strategy~\cite{yasin_DATE_2016}.} The chips are then unlocked, and thus, become functional.

\subsection{The SAT attack}
\begin{figure}[!t]
	\centering
	\includegraphics[width=0.35\textwidth]{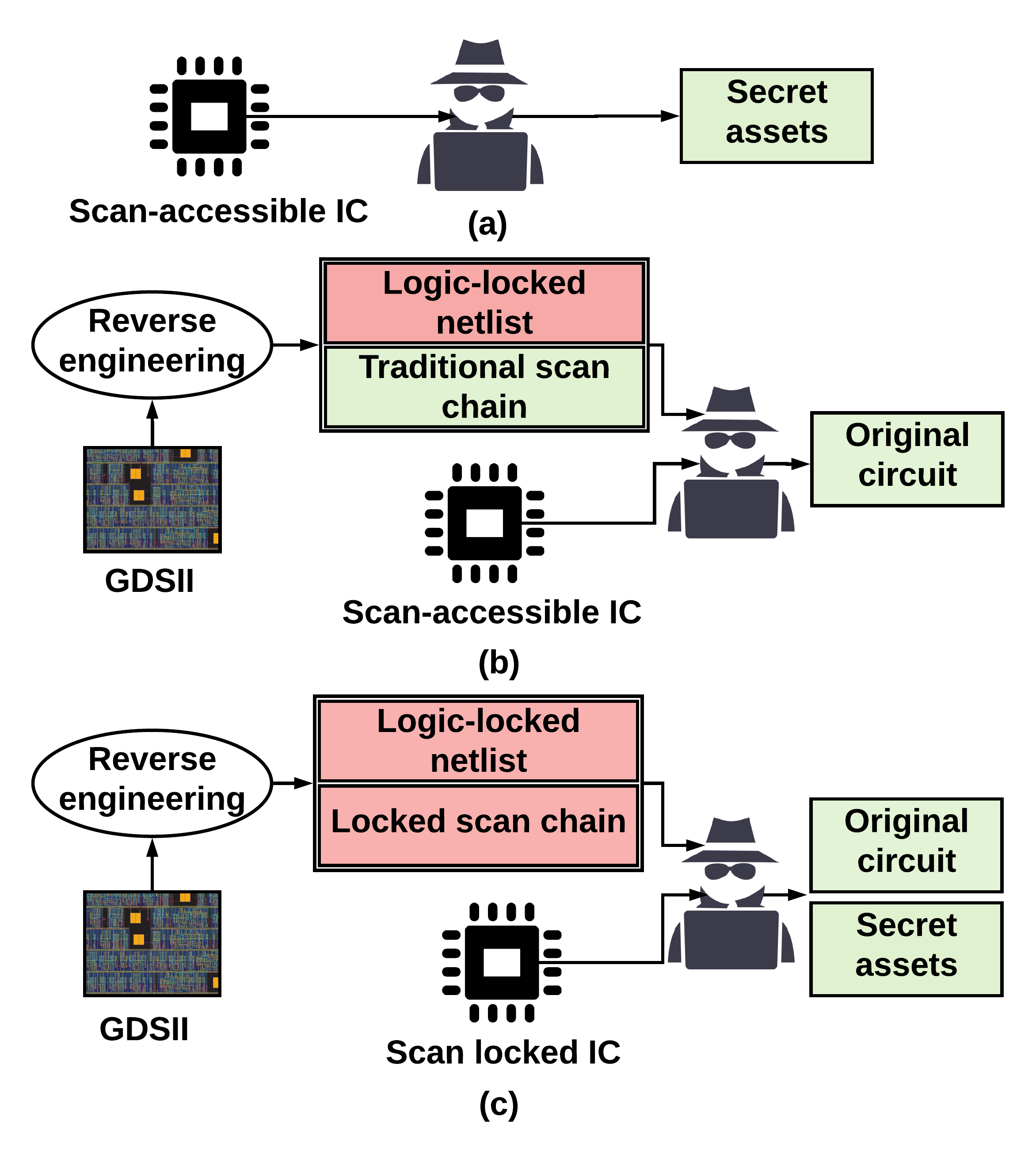}
	\caption{Threat models: (a) Scan-based attacks (b) SAT attack (c) ScanSAT attack}
	\label{fig:ScanSAT_threat_model}
\end{figure}

Logic and scan locking techniques are complementary defenses that protect the design IP and the test interface, respectively. A representative logic locking attack that aims at extracting the secret key is the so-called SAT attack~\cite{Subramanyan_host_2015}. The threat model showing the attacker's capabilities in the case of the SAT attack is displayed in Fig.~\ref{fig:ScanSAT_threat_model} (b). The attack requires a locked netlist (obtained through reverse engineering a chip or the layout information) and a working chip (obtained from the market) that is used as an oracle. The SAT attack applies a SAT solver on the Conjunctive Normal Form (CNF) representation of the locked netlist to produce a Distinguishing Input Pattern (DIP), which is an input combination for which at least two different key values generate differing outputs. The attack then applies this pattern to the working chip to obtain the correct response, which helps prune all the incorrect keys that fail to produce this output on the locked netlist. This process is repeated iteratively and the generated input-output pairs are added to the gradually growing CNF formulation. The attack succeeds when a DIP can no longer be found by the SAT solver, which is when the correct key is returned.

As the SAT attack broke all the logic locking techniques existing then, new logic locking techniques are recently proposed that are resilient to the attack. The new logic locking techniques can thwart the SAT attack by increasing the number of required DIPs exponentially. One recent solution is the Stripped Functionality Logic Locking (SFLL) technique~\cite{yasin_CCS_2017}. Yet, researchers are now developing functional analysis based attacks that can break certain instances of the SFLL technique~\cite{date2019_sirone,sfll_break,alrahis2019functional}.

What is common between the SAT attack and the scan-based attacks is that in both cases a full scan access is required. However, the threat model and the goal of the attacks are different. The threat model assumed for scan-based attacks is weaker than the one considered for the SAT attack and is shown in Fig.~\ref{fig:ScanSAT_threat_model} (a). The attacker is assumed to have access to the scan-accessible IC with assets inside (serial ID, crypto keys, IP details, etc.). The attacker can switch between the functional and test modes and steal the secret assets. 

Though the primary objective of scan locking is to thwart scan-based attacks that leak security-critical assets on chip, scan locking can be considered to presumably protect logic locking against SAT or other attacks as well.\footnote{We acknowledge that not all prior work on scan obfuscation claim such an objective. We carry out our security analysis in any case in order to clearly outline the limitations of scan locking.} Most of the logic locking attacks, including the powerful SAT attack, assume {\em full scan access} to a working oracle. We note that there are a few exceptions, such as~\cite{massad17}, that launch the attack without scan access; however, as explained further later, the effectiveness of such techniques is quite limited and they don't scale well. Scan locking thwarts the direct application of logic locking attacks as shown in Fig.~\ref{fig:threat_model}. 

\subsection{Contributions of This Work}
{\bf Powerful attacks such as SAT or scan attacks cannot break scan locking}, as scan access is now controlled with a secret key. \color{black} This is also argued in~\cite{wang2017secure}, which presents a dynamic scan locking solution that delivers protection against IP design piracy throughout supply chain in the presence of attacks such as the SAT attack. \color{black} In fact, as stated earlier, scan locking can be paired with logic locking to protect logic locking from any attack that requires scan access (i.e., SAT attack).

In this paper, we take on the challenge of breaking statically and dynamically obfuscated scan chains; we propose ScanSAT attack. We use Encrypt Flip-Flop (EFF)~\cite{karmakar2018encrypt} as a representative example for static scan obfuscation and the technique in~\cite{wang2017secure} as a representative example for Dynamically Obfuscated Scan (DOS), while we note that our attack can be adapted for other variants of scan obfuscation as well;  \color{black} we also show that we can break scan scrambling~\cite{hely2004scan} by adapting our ScanSAT modeling. \color{black}The attack flow is presented in Fig.~\ref{fig:threat_model} and the threat model is presented in Fig.~\ref{fig:ScanSAT_threat_model} (c). Consistent with almost all attacks on logic locking, the proposed attack requires (i) a working chip (with obfuscated scan chain(s)) and (ii) a locked design netlist. ScanSAT models the obfuscated scan chain as a logic locking problem, and then launches the SAT attack on it. It creates the combinational circuit equivalent of the scan-obfuscated circuit; this circuit is a logic-locked circuit with a key corresponding to the secret transformations on the scan chain(s). We then apply the SAT attack on this logic circuit model to extract the key; only then the SAT attack works successfully, simply because basic scan locking techniques do not account for the SAT attack when they embed transformations on the scan path.

The contributions of this work are as follows:
\begin{itemize}

    \item We show that our modeling can turn seemingly complex scan transformations into just another layer of breakable structures, helping break a unified scan and logic locking framework seamlessly. As opposed to regular combinational logic obfuscation, which is a 1-cycle transformation that depends on a key, scan obfuscation is an $n$-cycle transformation as scan operations are performed over $n$ cycles ($n$ is the scan depth). Scan obfuscation techniques have thus far relied on this difference in mitigating SAT attacks, which only apply to 1-cycle transformations. We show in this paper an attack that can handle an $n$-cycle transformation through a proper modeling and turn it into a 1-cycle transformation. 
    
    \item One important, yet counter-intuitive, finding is that simple static/dynamic scan locking falls short of protecting the scan access, and thus cannot be relied on in thwarting powerful attacks that need scan access to an oracle.
    \item We propose a novel attack ScanSAT that can break static and dynamic scan locking by extracting the scan locking key within a few iterations and a 100\% success rate.
    \item We show that ScanSAT is effective even for large key sizes and in the presence of on-chip scan compression.
    \item We show that ScanSAT can also break a two-layered defense where scan locking is coupled with a SAT-attack-vulnerable logic locking technique; we show that scan locking fails to protect logic locking against the SAT attack. A SAT attack resilient logic locking solution is still required to protect the design IP.
\end{itemize}

\color{black} ScanSAT has not been designed to be effective on other protection solutions that are outside of the scan obfuscation category; naturally, ScanSAT cannot break them. These defenses are partial scan, scan pin defusion, and those that involve test mode control and scan chain masking, such as the Design-for-Security (DFS) architecture proposed in~\cite{guin2018robust}, which recently got broken by a custom attack to be presented in ICCAD 2019.

\begin{figure*}[!t]
	\centering
	\includegraphics[width=0.9\textwidth]{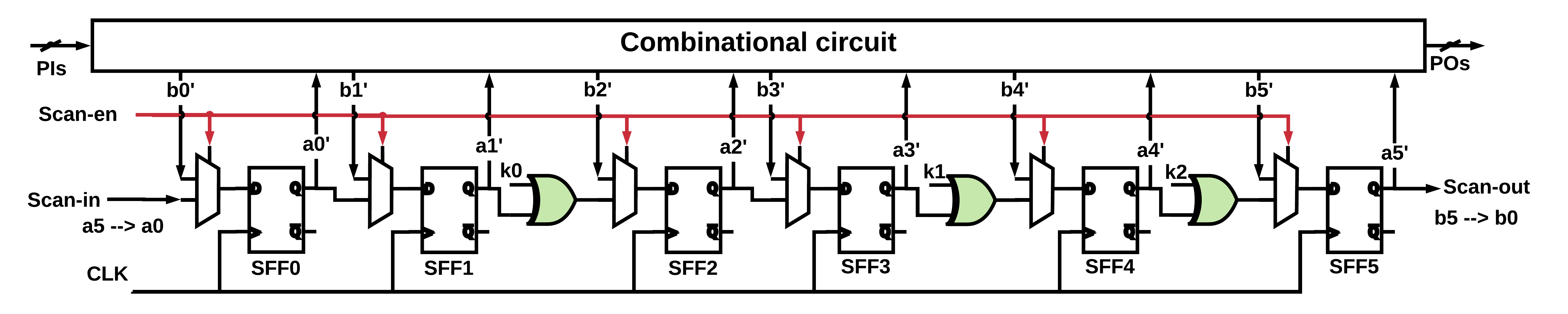}
	\caption{Scan obfuscation on the $s386$ benchmark; three key-bits obfuscate the scan operations.}
	\label{fig:encrupt_scan_chain_xor}
\end{figure*}

\color{black}
\section{Static Scan Obfuscation}

\subsection{Idea and Implementation}
The idea is to obfuscate the scan path via secret inversions; this way, the patterns delivered to the scan chain(s) differ from the scan-in stimulus in a secret manner. In addition, the captured response differs from the scan-out pattern that is observed through the Scan-out pin(s). A secret key dictates these inversion operations on the scan path, and thus, the exact relationship between: (i) the scan-in and delivered patterns and (ii) the captured and scan-out patterns.

Secret inversions can be inserted into the scan chains by inserting XOR gates (in alternative implementations, MUXes) between selected scan cells. 
An XOR gate driven by a key-bit of 1 implements an inversion. A designer can insert $k$ XOR gates, resulting in a $k$-bit key; some of these $k$ XORs implement inversion on the scan chain. Although a reverse engineer can identify the locations of these gates on the scan path, only the designer knows the key, and thus, the subset of gates that perform an inversion operation.

To create a search space for the attacker that is exponential in the size of the key, naive scan obfuscation can initially embed scan inversions in selected locations, some of which are canceled by XORs driven by a key-bit of 1; this way, the attacker cannot make any deductions about the key value from the location of the XOR gates. All commercial Automatic Test Pattern Generation (ATPG) tools are able to handle scan inversions; the designer with the secret key information can perform test generation in the presence of these inversions.

\subsection{Obfuscated Scan Chain Example}

An example obfuscated scan chain is presented in Fig.~\ref{fig:encrupt_scan_chain_xor} for the circuit $s386$ from the ISCAS-89 benchmark suite~\cite{iscas89}. Three key-bits $k_{0}, k_{1}$, and $k_{2}$ obfuscate the scan operations. The scan-in pattern applied from the Scan-in pin is denoted by $a$, where $a_i$ is the bit intended for $SFF_i$.

In this example, let's assume that the secret key is $111$ and thus all three XORs insert inversion in a secret fashion. The pattern delivered into the scan cells upon the completion of six shift cycles is denoted by $a^{'}$, where $a_{i}^{'}$ is the bit delivered into $SFF_i$. Due to the inversions introduced by the locking XORs, $a \ne a^{'}$. In this example, $SFF_2$, $SFF_3$, and $SFF_5$ receive their stimuli inverted due to an odd number of inversions between the Scan-in pin and these flip-flops, while $SFF_0$, $SFF_1$, and $SFF_4$ receive their stimuli as is due to an even number of inversions. 

During the capture operation, $a^{'}$ is applied to the combinational circuit rather than $a$. After the capture cycle, the SFFs will capture their corresponding functional inputs (response of the combinational circuit) denoted by $b^{'}$. However, the same locking XORs apply inversions on the response bits as well; the pattern observed through the Scan-out pin is denoted as $b$, where $b_i$ corresponds to $SFF_i$. Due to the inversions introduced by the locking XORs, $b \ne b^{'}$. In this example where all XORs insert inversion, the captured response bits in $SFF_2$, $SFF_3$, and $SFF_5$ are observed as is (even number of inversions from the flip-flop to the Scan-out pin), while response bits in $SFF_0$, $SFF_1$, and $SFF_4$ are inverted prior to being observed through the Scan-out pin (odd number of inversions). A test generation (ATPG) tool with the knowledge of the secret key would be aware of the scan inversions, thus producing transformed stimuli $a$ and responses $b$ automatically.

\subsection{Security Claims}

The ever-assumed equivalence of scanned-in to delivered stimuli and captured to observed responses is broken by scan locking in a secret manner; $a \ne a^{'}$ and $b \ne b^{'}$. The secret key dictates the transformation from $a$ to $a^{'}$ and the one from $b$ to $b^{'}$. An untrusted party is able to use $a$ and $b$ to test chips through the scan interface, yet without being able to infer the security-critical data $a^{'}$ and $b^{'}$, effectively thwarting attacks that rely on scan access. 

Simple scan-flush attempts by an attacker where special patterns such as all 0's or all 1's are shifted through scan chain(s) with no capture operation, reveal very limited information about the secret inversions on the scan path. Such attempts only reveal whether the total number of inversions between the Scan-in pin and the Scan-out pin of the entire chain(s) is even or odd. Where exactly on the scan chain(s) these inversions take place remains to be a mystery for the attacker.

\subsection{Scan Chain Scrambling}
Another way to transform scan data is through scan chain scrambling, in which the order of SFFs in the scan chains is hidden and controlled by a secret key~\cite{hely2004scan}. In scan chain scrambling, the scanned-in/scanned-out data is permuted. In order to scramble the scan chains, MUXes are added between selected SFFs, where the scan input of the selected $SFF_{j,i}$ (located on scan chain $j$, scan slice $i$) is fed by the output of a MUX. The inputs to the MUX can come from any two (or more) SFFs from the $(i-1)^{th}$ scan slice. The select lines of the newly added MUXes are secret key-bits that control the ordering of scan path fragments.

\begin{figure}[!t]
	\centering
	\includegraphics[width=0.4\textwidth,trim=3in 3in 3in 2in]{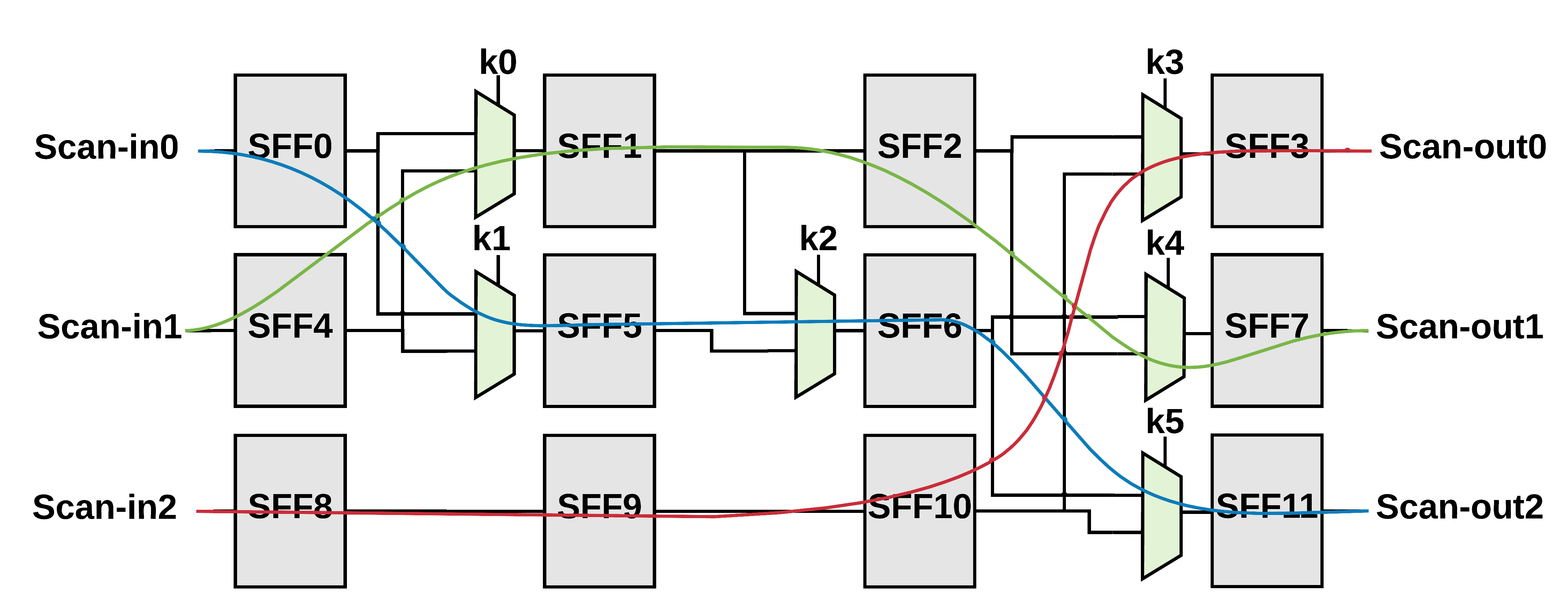}
	\caption{Scan chain scrambling implementation; six key-bits obfuscate the scan path. The correct key is 101110 and the correct (secret) scan paths are colored in blue, green and red.}
	\label{fig:scramble}
\end{figure}

Figure~\ref{fig:scramble} presents an example of scan chain scrambling implementation. Six MUXes are added between the scan elements. The three bolded lines in blue, green and red demonstrate the three correct scan paths; the correct key is 101110. Without the knowledge of the secret key, the correct scan path is hidden from the attacker. Therefore, the end goal of scan obfuscation is achieved; scan data is secretly transformed (permuted, in this case).

\begin{figure}[!t]
	\centering
	\includegraphics[width=0.4\textwidth,trim=3in 3in 3in 2in]{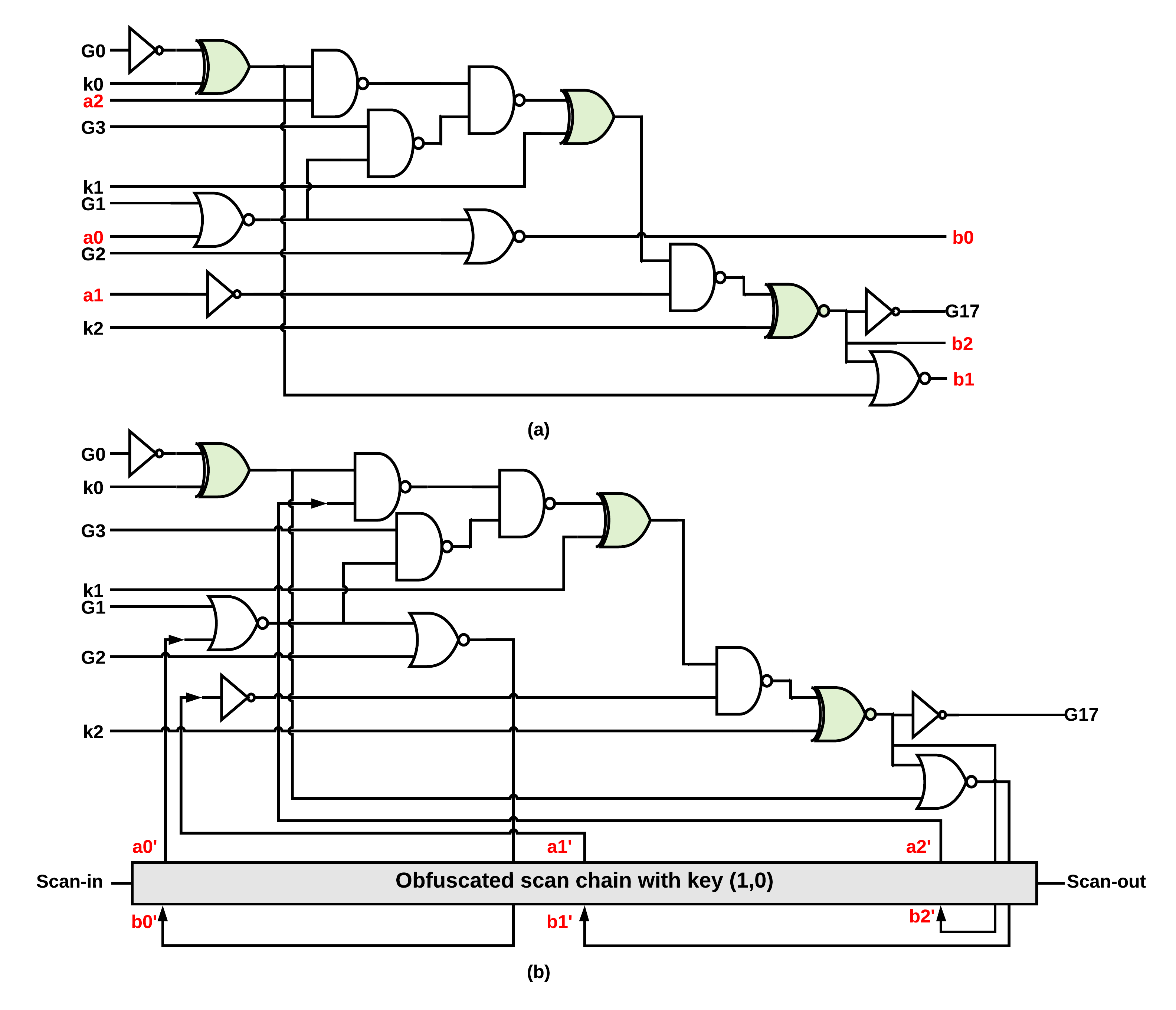}
	\caption{(a) The $s27$ benchmark locked using XOR/XNOR gates; three key-bits $k_{0}\longrightarrow k_{2}$ lock the combinational circuit. The correct key is 101. The inputs/outputs of the flip flops are converted into PPIs/PPOs marked in red. (b) Adding scan obfuscation as an additional layer. Scan obfuscation blocks the access to the scan chain and thus the inputs/outputs of the flip flops are no longer accessible.}
	\label{fig:locked_netlist}
\end{figure}

\subsection{SAT Attack on Static Scan Obfuscation}

We locked the $s27$ circuit from the ISCAS-89 benchmark suite~\cite{iscas89}, containing three flip-flops, by using three key-gates as shown in Fig.~\ref{fig:locked_netlist} (a). Since $s27$ is a sequential circuit, the inputs ($a_{0}$, $a_{1}$, and $a_{2}$) and outputs ($b_{0}$, $b_{1}$, and $b_{2}$) of the FFs are turned into Pseudo Primary Inputs (PPI)s and Pseudo Primary Outputs (PPO)s. We then ran the SAT attack; it was successful in retrieving the secret key value, \color{black} as the attack was able to handle the 1-cycle transformation implemented by logic locking. \color{black}

The existing static scan locking techniques are claimed to be resilient against the SAT attack due to blockage of the scan access. To verify this claim, we added scan obfuscation as an additional security layer as shown in Fig.~\ref{fig:locked_netlist} (b). Two additional key-bits obfuscate the scan operations with the pattern $(1,0)$ for the three-bit scan chain. The SAT attack assumes a full access to the scan chain, therefore the scan obfuscation is ignored and the locked circuit's netlist is assumed to be the one displayed in Fig.~\ref{fig:locked_netlist} (a). The goal of the SAT attack is to identify the logic locking key. Yet the SAT attack fails to obtain the correct key, since upon applying the obfuscated responses as constraints, all possible key values were eliminated from the search space.
\color{black} \textbf{The SAT attack was thus unsuccessful against the 3-cycle transformation implemented by static scan obfuscation and the SAT solver returned an UNSAT}.

\color{black}
\section{Dynamic Scan Obfuscation}
\label{sec:dynamic_scan_obfuscation}
\subsection{Idea and Implementation}
Similar to static scan obfuscation, dynamic scan obfuscation inserts secret inversions as well, obfuscating the scan path. However, in this case, the obfuscation key changes over time; specifically in~\cite{wang2017secure}, the obfuscation key changes with a key update frequency $p$. As a result, the test data transformation functions change over time, resulting in a stronger defense. From the attacker's perspective, there is now a sequence of keys to extract as opposed to a single key.

A Dynamically Obfuscated Scan (DOS) architecture was proposed in~\cite{wang2017secure}. The structure consists of (i) a Linear-Feedback Shift Register (LFSR) (ii) a shadow scan chain, and (iii) a control unit. This DOS structure is highlighted in grey rectangle in Fig.~\ref{fig:DOS}. 

The \textbf{LFSR} is used in order to generate a $\lambda$-bit obfuscation key where $\lambda$ is the maximum length of a scan chain. The LFSR operation is driven by the control unit and it will change its output when a key update signal is received.

The \textbf{shadow scan chain} consists of $\lambda$ flip-flops and it takes in the $\lambda$-bit obfuscation key generated by the LFSR. The outputs of the chain are $[x\times\lambda\times\alpha]$-bits, where $x$ is the number of scan chains in the DfT structure and $\alpha$ is the permutation percentage (the percentage of SFFs selected to be locked in each scan chain). The purpose of the shadow scan chain is to prevent reset/flush attacks that assume that all scan cells are reset upon power-up. Therefore, it aims to protect the obfuscation key from being leaked. Shadow chain makes sure that all 0's are read out from the Scan-out pin(s) during the shift-in of the first pattern after power-up. Once the scan chain(s) are filled with the first pattern, the shadow chain has no purpose.

The \textbf{control unit} consists of an $n$-bit register, an $n$-bit counter, and a control flip-flop. Upon startup, a secret control vector is loaded from a protected memory. The control vector consists of the $\lambda$-bit secret seed to be loaded into the LFSR, the $p$ value representing the key update frequency, and the maximum allowable number of key updates. This secret information is chosen by the designer.

When scan operation is enabled, the $n$-bit counter increments with each capture pulse. Once $p$ value is reached, a key update signal is generated, causing the LFSR to change its output and the counter to reset.

\subsection{Dynamic Scan Obfuscation Example}
\begin{figure}[!t]
	\centering
	\includegraphics[width=0.46\textwidth]{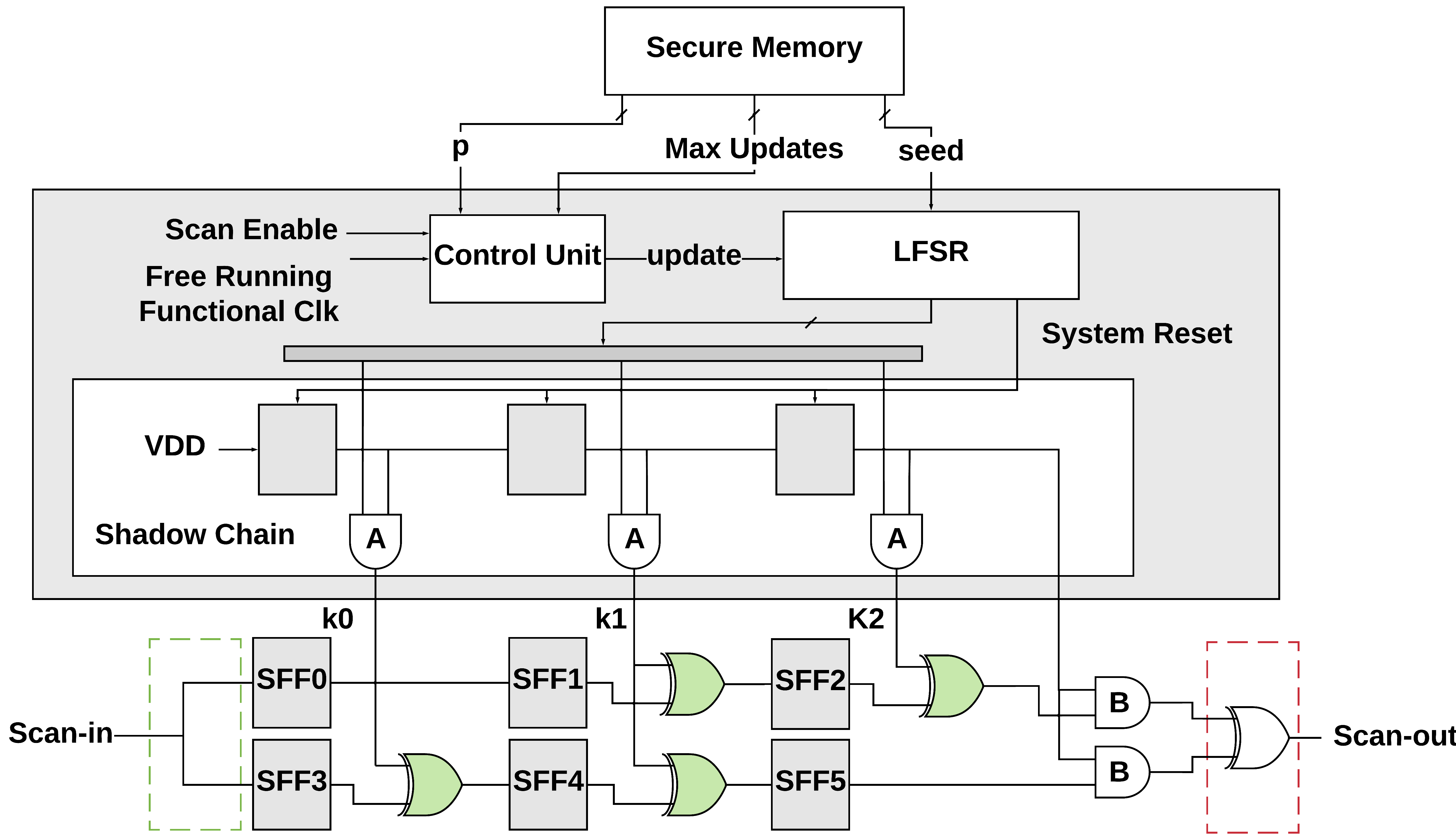}
	\caption{Dynamic scan obfuscation on the s386 benchmark; $\lambda=3$, $x=2$ and $\alpha=2/3$~\cite{wang2017secure}.}
	\label{fig:DOS}
\end{figure}

The s386 circuit is obfuscated using the DOS technique as shown in Fig.~\ref{fig:DOS}. The key size in this example is $\lambda=3$, and thus, a three-bit LFSR is utilized. Upon startup, the key update frequency $p$, the secret three-bit seed, and the maximum allowable number of key updates will be read from memory. A compression structure is present in this example comprising a fanout stimulus decompressor (highlighted in green dotted rectangle) and an XOR response compactor (highlighted in red dotted rectangle). The six SFFs available in the design are divided into two scan chains, i.e., $x=2$. Four XOR gates in total are inserted in the scan chains in order to provide scan locking similar to static scan obfuscation. The difference here is that the three-bit key $(k0,k1,k2)$ changes over time based on an update frequency $p$.

\subsection{Security Claims}
Dynamic scan obfuscation promises all the security guarantees (resilience to scan flush and mode-reset attempts, no key inference from XOR locations, etc.) of static scan obfuscation. In addition, resetting attacks will not work due to the utilization of the shadow scan chain which protects the obfuscation key during a global reset. The shadow scan chain also protects from differential attacks~\cite{rolt2013novel_old}, as the first scan-out pattern is all 0's. Furthermore, as the key is updated periodically, there is a limited time window for any attack to retrieve a key before it is updated.

\section{ScanSAT on Static Scan Obfuscation}
\begin{figure}[!t]
	\centering
	\includegraphics[width=0.45\textwidth]{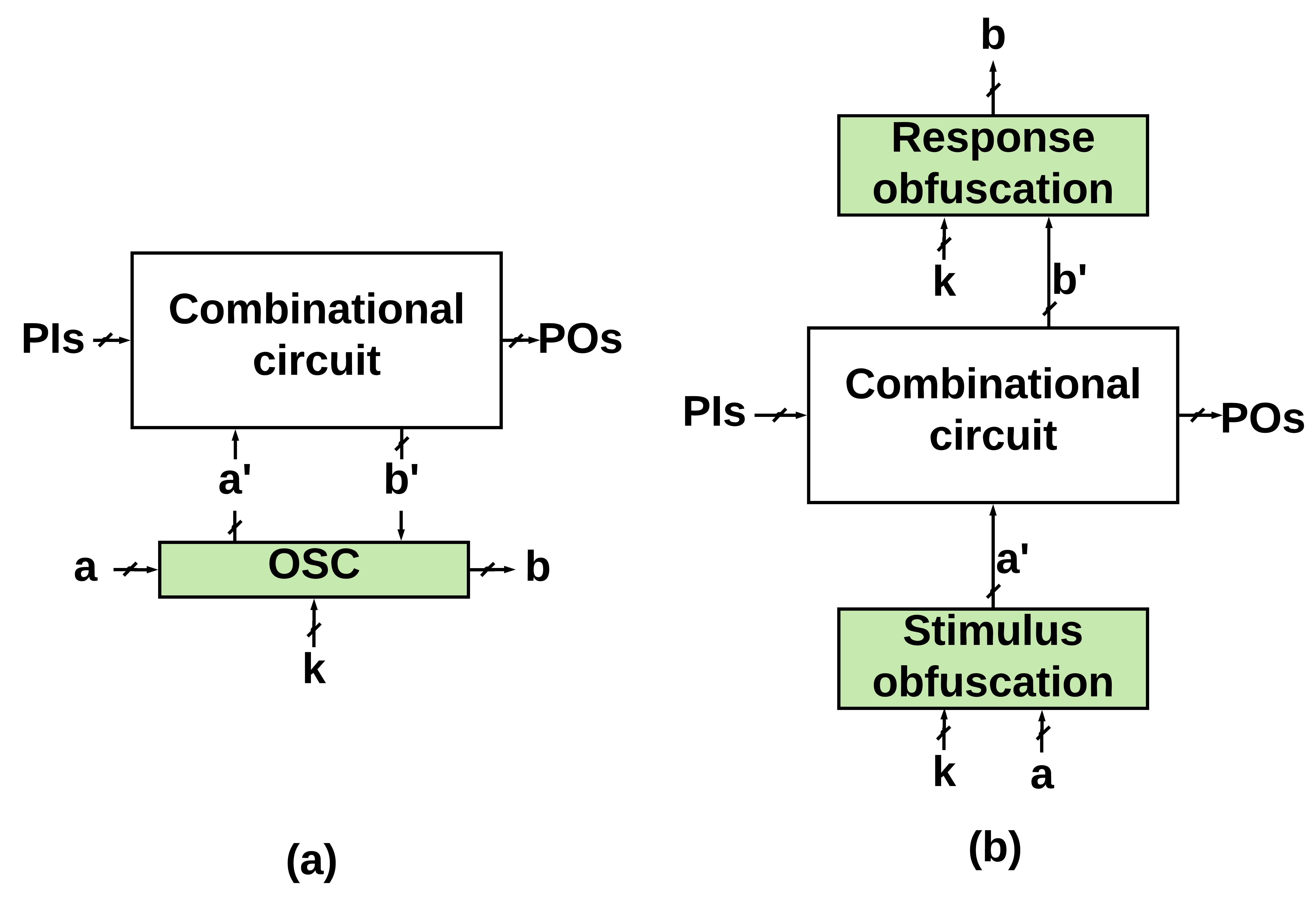}
	\caption{(a) Obfuscated scan chain. (b) Modeling the obfuscated scan chain(s) using ScanSAT.}
	\label{fig:ScanSAT_concept}
\end{figure}

In this section, we present the ScanSAT attack on static scan obfuscation techniques. 
ScanSAT is based on the insight that the complex $n$-cycle transformation of scan locking can be modeled by generating a combinational (1-cycle transformation) equivalent of the scan-obfuscated circuit. The obfuscation inversions on the scan path become part of the resultant combinational circuit, which effectively is a logic-locked circuit with key logic inserted at the pseudo-primary inputs/outputs of the circuit. The logic-locked circuit equivalent of a generic scan-obfuscated circuit in Fig.~\ref{fig:ScanSAT_concept}(a) is provided in Fig.~\ref{fig:ScanSAT_concept}(b), where the obfuscation on the stimulus and the response are modeled separately as combinational blocks driven by the same scan obfuscation key. The resultant circuit can now be attacked via traditional logic locking attack techniques, such as the SAT attack developed by Subramanyan et al. \cite{Subramanyan_host_2015} or the test-data mining attack formulation developed by Yasin et al.~\cite{yasin_DATE_2016}. 
Breaking the logic-locked circuit and extracting its key is equivalent to breaking basic scan obfuscation and deobfuscating/unlocking the scan chain(s). 

The proposed ScanSAT modeling is also capable of accounting for any additional logic locking technique applied in conjunction to scan obfuscation; the circuit in Fig.~\ref{fig:ScanSAT_concept}(b) allows for the incorporation of another logic locking technique (with a separate key) applied on the combinational circuit. In that case, the resultant logic-locked circuit that models two layers of defenses can then be attacked via the SAT attack, extracting both keys simultaneously. We revisit this point in section~\ref{sec:RLL}.

\subsection{ScanSAT on Key-gate Insertion Based Obfuscation}
ScanSAT comprises two basic stages: (i) modeling the obfuscated scan chain as a logic locking problem and (ii) breaking the obtained modeled circuit using attacks on logic locking.

The first step is the formulation of the relationship between scan-in pattern $a$ and the pattern delivered into the scan chain(s) $a^{'}$. Some of the bits in $a^{'}$ are identical to the corresponding bits in $a$ while the remaining ones are complementary to the corresponding bits in $a$. The secret key value $k$ dictates the exact relationship; the bits that pass through an even number of inversions are delivered intact, while those that pass through an odd number of inversions between the Scan-in pin and the target scan cell are inverted. In the example in Fig.~\ref{fig:encrupt_scan_chain_xor}, the following equations capture the relationship between $a$ and $a^{'}$:

\begin{equation}
a_{i}^{'}= a_{i} \oplus L(k,i) 
\label{eq:ai}
\end{equation}

\noindent where $L(k,i)$ is XOR of all the key bits of key $k$ that are injected into a location between the Scan-in pin and scan cell $i$. Without knowing the value of $k$, the attacker cannot tell which stimulus bit is delivered intact and which one is inverted. 

\begin{figure}[!t]
	\centering
	\includegraphics[width=0.45\textwidth]{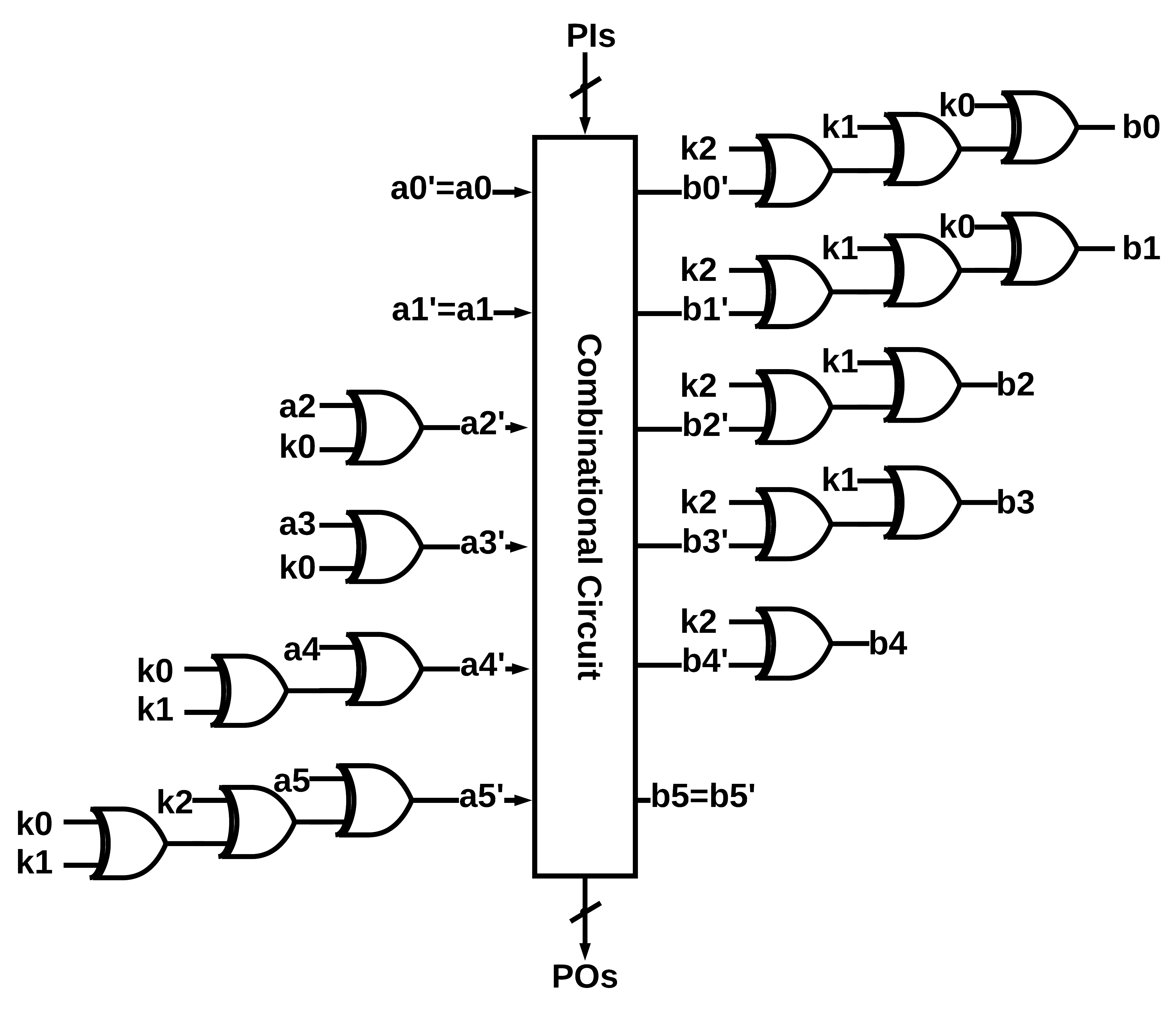}
	\caption{Modeling the obfuscated scan chain(s) as a logic locking problem.}
	\label{fig:model_example}
\end{figure}

A capture operation produces the response $b^{'}$ upon applying the scan chain(s) content $a^{'}$ to the combinational circuit. The other obfuscation layer consists of the unknown number of inversions each captured response bit passes through prior to being observed on the Scan-out pin. Again, the values of  $k$ dictate these secret inversions. The relationship between the captured response pattern $b^{'}$ and the observed scan-out pattern $b$ can be formulated similarly as follows:
\begin{equation}
b_{i}= b_{i}^{'} \oplus R(k,i) 
\label{eq:bi}
\end{equation}
\noindent where $R(k,i)$ is XOR of all the key bits of key $k$ that are injected into a location between the scan cell $i$ and the Scan-out pin. The equations above can be easily modeled as additional XOR logic around the combinational circuit, relating $a^{'}$ to $a$ and $b$ to $b^{'}$ as a function of the secret key $k$. For the same example, the modeled circuit is shown in Fig.~\ref{fig:model_example}. This modeled circuit captures the transformation of inserted scan-in pattern $a$ to the pattern delivered in scan chain(s) $a^{'}$, which is applied through the pseudo primary inputs of the combinational circuit. It also captures the transformation of the captured response pattern $b^{'}$ to the scan-out pattern $b$. On the input side, the same key-bit/gate appears multiple times as it affects all the flip-flops to its right on the scan path. Similarly, on the output side, the same key-bit/gate appears multiple times as it affects all the flip-flops to its left on the scan path. This modeling can also be conceived as unrolling of the scan operations. 

The final modeled circuit in Fig.~\ref{fig:model_example} is simply a logic-locked circuit that has three key-bits. An attacker can use this modeled circuit along with the scan-obfuscated oracle to identify the secret key $k$. For this, the attacker runs the SAT attack \cite{Subramanyan_host_2015} on this modeled circuit, generating (obfuscated) input-output patterns.
The input patterns $a$ are those the attacker then applies from the Scan-in pin of a working chip. The output patterns $b$ are those the attacker collects from the Scan-out pin of the working chip. By iteratively generating the input-output patterns, the attacker gradually prunes the key search space, and produces the secret key $k$ of the logic-locked circuit, which is also the key used to obfuscate the scan chains.

\subsection{ScanSAT on Scan Chain Scrambling}
In this section, we explain how ScanSAT can be adapted to break scan chain scrambling; the ScanSAT modeling is tweaked to account for the scrambled scan operations. In the case of scan chain scrambling, multiplexer logic that captures the scan chain reordering operations becomes part of the modeled combinational circuit. The first step in generating the logic-locked circuit equivalent is the formulation of the relationship between the scanned-in pattern $a$ and the pattern delivered into the scan chain(s) $a^{'}$. Next is the formulation of the relationship between the captured response $b^{'}$ and the scanned-out pattern $b$. The same scrambling key affects both the stimulus and response transformations.

\begin{figure}[!t]
	\centering
	\includegraphics[width=0.45\textwidth]{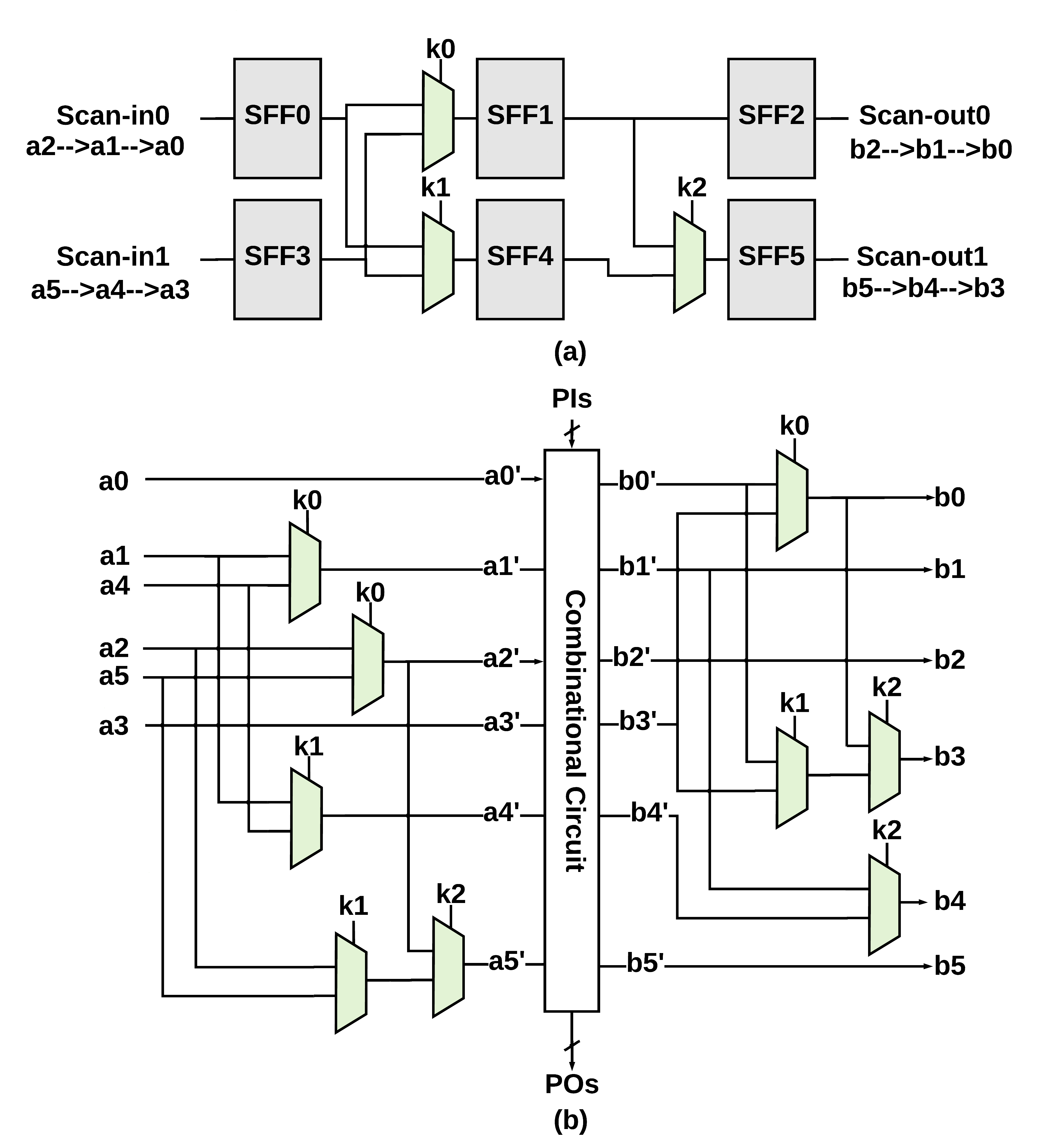}
	\caption{(a) Scan chain scrambling; three key-bits are used. (b) Modeling as a logic locking problem.}
	\label{fig:scansat_scramble}
\end{figure}

The difference between key insertion based scan obfuscation and scan scrambling is that in the latter case, the transformations between $a$ and $a^{'}$ and between $b^{'}$ and $b$ are not one-to-one mappings. The bit delivered into $SFF_{i,j}$ (SFF located on scan chain $j$, scan slice $i$) can come from any Scan-in$_{j}$ input and the captured bit can pass through any Scan-out$_{j}$ output, depending on the secret key and the location of the MUXes.
The scrambling implementation example using three key-bits shown in Fig.~\ref{fig:scansat_scramble} (a) is modeled for attack in the form of the logic-locked circuit in Fig.~\ref{fig:scansat_scramble} (b). SAT attack is executed on Fig.~\ref{fig:scansat_scramble} (b) to retrieve the secret key, which reveals the secret shuffling of the scan chain fragments.

\section{ScanSAT on Dynamic Scan Obfuscation}
Next, we discuss how ScanSAT can be extended in order to break dynamic scan obfuscation. As explained earlier, dynamic scan obfuscation uses a sequence of keys generated by an LFSR based on a secret seed and a key update frequency $p$, which is also a secret. The fact that the keys change in every $p$ patterns necessitates a radically different approach than simply modeling the scan transformations and running the SAT attack, which worked well for static scan locking as we have shown, but falls short in breaking dynamic scan locking. SAT attack has been developed for logic locking defenses that depend on a static key that remains constant throughout the lifetime of a chip; an attack that extracts dynamic keys is missing in logic locking literature.

While the seed and $p$ are secret, both the LFSR structure (from the reverse-engineered netlist) and its polynomial are available to the attacker. The first objective of ScanSAT is to identify the value of $p$. This value is then used to identify the secret seed. From the secret seed and $p$, the attacker can derive all the keys that are dynamically generated on-chip. ScanSAT on dynamic scan obfuscation comprises the following steps.

\subsection{Step 1 - Identify the key update frequency $p$} 

A simple approach to identify $p$ is to apply the same stimulus pattern $Stim$ repeatedly from the Scan-in pins, and observe the response $Resp$ through the Scan-out pins; $Stim$ transforms into $Stim'$, which is applied to the combinational logic, and the captured response $Resp'$ is transformed into $Resp$. Upon $p$ capture operations, the key is updated. From that point on, when the same $Stim$ is applied, it is now transformed to $Stim''$ and applied to the combinational circuit; the captured response is $Resp''$. $Resp''$ now goes through a different transformation than $Resp'$ because of the new key; therefore, most likely, there will be a noticeable change in the observed response, helping detect the obfuscation key update operation:

\begin{equation}
Resp_j = R(C(L(Stim,k_j)),k_j)
\label{eq:identify_p}
\end{equation}

In the equation above, $L()$ and $R()$ are the linear transformation operations that the stimulus and the response, respectively, go through due to scan obfuscation with the $j^{th}$ dynamic key $k_j$; these operations were already defined in the previous section. $C()$ is the function implemented by the combinational logic. Even though the same stimulus $Stim$ is applied, the $j^{th}$ response $Resp_j$ (with dynamic key $k_j$ active) will likely be different from $Resp_{j-1}$ (with dynamic key $k_{j-1}$ active), as $k_j$ is different from $k_{j-1}$.

\textbf{Challenge:} It is theoretically possible, though unlikely, that $Resp_j$ and $Resp_{j-1}$ are identical. This happens when the stimulus transformation $L()$, the capture operation through the combinational logic function $C()$, and the response transformation $R()$ collectively produce identical responses even for two different keys $k_j$ and $k_{j-1}$. This is possible because although $L()$ and $R()$ are linear functions, $C()$ is not. In this highly unlikely scenario, the key update operation would go unnoticed. 

\textbf{Solution:} A simple remedy is to repeat the same process with different stimulus patterns. The smallest one of the identified values of $p$ can be considered as the actual $p$. The more stimulus patterns used in the process, the higher the confidence level. In our experiments, we never encountered this unlikely scenario.

\subsection{Step 2 - Extract the secret seed} 

As opposed to the static scan locking case, we can't run the SAT attack the same way, as a key remains valid for only $p$ patterns. The SAT attack normally generates all the DIPs aiming for one static key, which is no longer the case in dynamic scan locking. From Step 1, we now know $p$, and thus, the number of patterns for which a key remains valid. 

Similar to the ScanSAT on static scan locking, we model the $L()$ and the $R()$ functions in the form of XOR circuitry inserted around the combinational logic for dynamic scan locking as well. The difference now is that the SAT attack can be executed for at most $p$ patterns, which defines the time window for which each dynamic key remains valid. If more than $p$ DIPs are required to identify a dynamic key, the SAT attack needs to be terminated prematurely upon $p$ DIPs. Another SAT attack can be executed subsequently to identify the next dynamic key in the sequence; this attack too will have to be terminated upon $p$ DIPs. The question is how to combine all the information obtained from these prematurely terminated independent SAT attack runs in identifying the secret seed.

\begin{figure}[!t]
	\centering
	\includegraphics[width=0.37\textwidth]{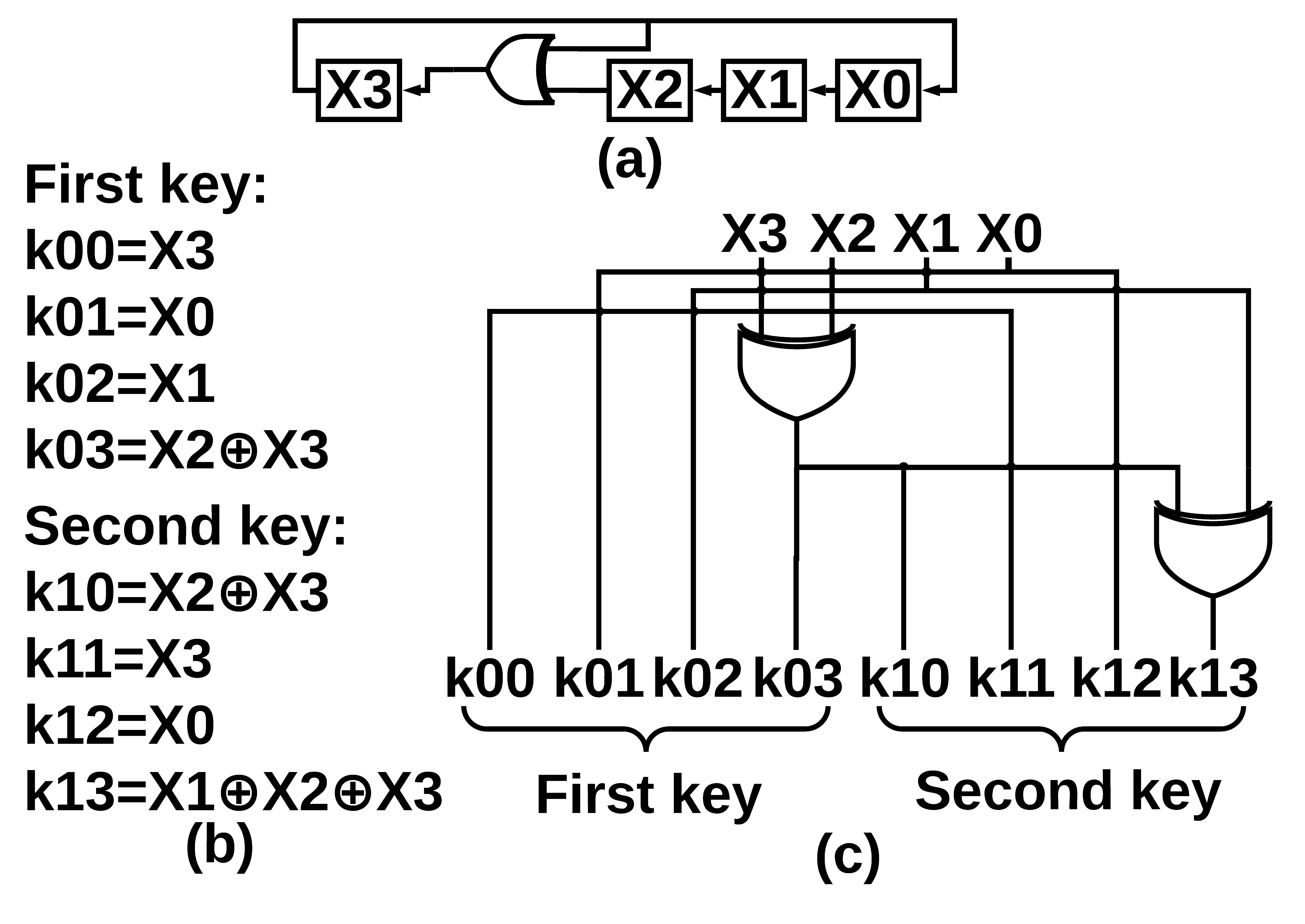}
	\caption{(a) LFSR structure, (b) equations that relate the first two four-bit keys $k_{ij}$ to the seed $X_i$, and (c) Seed-to-key block for the first two keys.}
	\label{fig:LFSR_model}
\end{figure}
\begin{figure}[!t]
	\centering
	\includegraphics[width=0.35\textwidth]{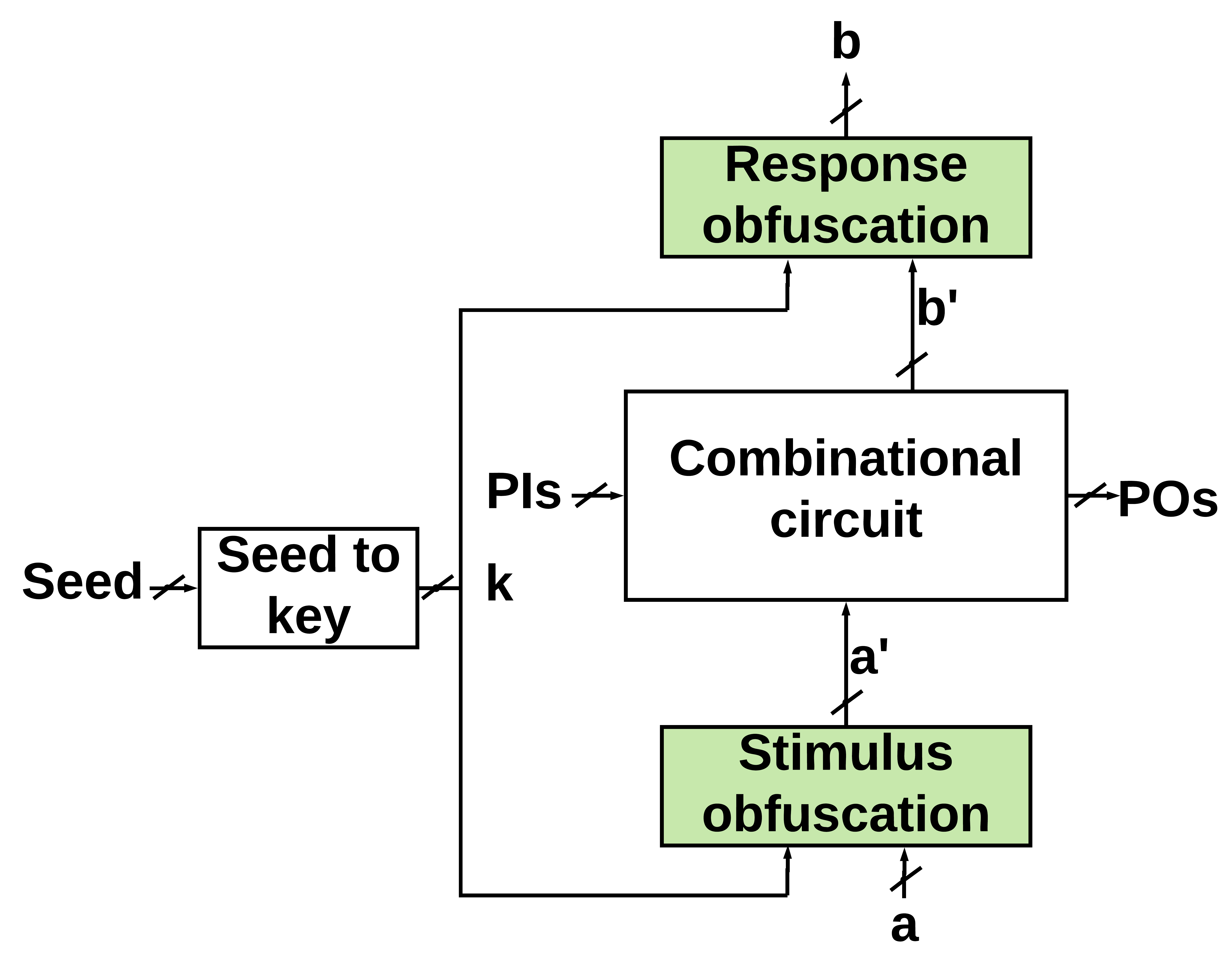}
	\caption{ScanSAT modeling to compute the seed.}
	\label{fig:seed-to-key}
\end{figure}

All the dynamic keys are generated from one secret seed based on the LFSR structure (primitive polynomial). A set of linear equations, known to the attacker, define the relationship between the secret seed and each dynamic key $k_j$. The example in Fig.~\ref{fig:LFSR_model} shows the linear equations mapping the four-bit seed to the first two four-bit dynamic keys generated from this seed. Although the independent SAT attack runs aim for identifying the dynamic keys, they in fact reveal information about the seed every time a dynamic key becomes partially known. It is possible to combine information from independent SAT attack runs by gradually gathering information about the seed in every run.

We can therefore incorporate into our ScanSAT model the relationship between the seed and the keys. As shown in Fig.~\ref{fig:LFSR_model}, an LFSR can also be modeled as a combinational block consisting of XOR gates; we refer to this block as the {\em seed-to-key} block. We append an instance of this block into the logic-locked circuit equivalent of scan-obfuscated design as in Fig.~\ref{fig:seed-to-key}, forcing ScanSAT to solve directly for the seed. However, when the SAT attack is terminated upon $p$ DIPs prematurely, the seed is resolved only partially. Fortunately, SAT attack run prematurely terminates with a Conjunctive Normal Form (CNF) formula that captures the information on the partially resolved seed thus far, which can be carried over to the next SAT attack run (with the next dynamic key in the sequence). During the next SAT attack run, the seed-to-key block needs to be updated based on the linear equations that define the relationship between the seed and the new dynamic key, and the new SAT attack run is executed by starting with the CNF from the previous run(s). 

\begin{eqnarray}
a_{i}^{'} &=& a_{i} \oplus L(S(seed, j),i)  \\ 
b_{i} &=& b_{i}^{'} \oplus R(S(seed, j),i) 
\label{eq:CNFs}
\end{eqnarray}

\begin{figure}[!t]
	\centering
	\includegraphics[width=0.15\textwidth]{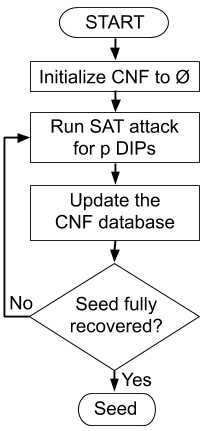}
	\caption{Iterative execution of ScanSAT on dynamic scan locking by using independent SAT attack runs.}
	\label{fig:flowchart}
\end{figure}

The CNF of the $j^{th}$ SAT attack run $CNF_j$ is formed by running the SAT attack for up to $p$ DIPs based on the modeling captured by the equations above; $a_i$, $b_i$, $a_{i}^{'}$, and $b_{i}^{'}$ were already defined for ScanSAT on static scan locking. $S()$ function is implemented by the seed-to-key block; $S(seed, j)$ denotes the $j^{th}$ dynamic key obtained from the seed. $CNF_j$ is the input to the $j+1^{th}$ SAT attack run to produce $CNF_{j+1}$. Our attack terminates when one of the SAT attack runs fully resolve the seed with the help of the CNFs from the prior runs as shown in the flowchart in Fig.~\ref{fig:flowchart}.

The number of independent SAT attack runs required to resolve the secret seed depends on the value of $p$. For large values of $p$, a few SAT attack runs with a large number of DIPs will be able to resolve the seed, while a small value of $p$ (frequent key updates) will result in a large number of SAT attack runs each terminating prematurely with a small number of DIPs. For $t$ independent SAT attack runs (or alternatively, $t$ iterations), ScanSAT ends up generating around $t \times p$ DIPs; each iteration except for the last one terminates with $p$ DIPs, while the last iteration may terminate with fewer than $p$ DIPs. Collectively, the seed will eventually be identified as the proposed attack is capable of carrying partial information about the seed from one independent SAT attack run to the next. Once the seed is identified, the entire key sequence can be reproduced based on the polynomial of the LFSR and the $p$ value.

\subsection{Example}

We now explain the working of our attack on a small example for the most challenging case of $p=1$ in the dynamic scan obfuscation scheme in~\cite{wang2017secure}; the dynamic key is updated in every capture cycle, resulting in a different key for every test pattern, and thus, necessitating that each independent SAT attack run be executed for only one DIP.

Consider a sequential circuit whose scan chain is locked with 5 key bits. The five-bit key is dynamically updated for every pattern based on the seed 00001 loaded on a five-bit LFSR. Table~\ref{tab:dyn_lock} shows the correct values of the dynamic keys for four cycles along with how many of these key bits were recovered in each SAT attack run. The first SAT attack run, for example, is terminated with a single DIP for the first key; at that time, two bits of the first key are identified to be 0's. The same table also reports the information revealed about the seed in each cycle; the two recovered bits of the first dynamic key helps reveal two bits of the seed by solving the linear equations obtained from the LFSR structure. The next SAT attack run on the second pattern (and thus, the second key) helps identify four bits of the second key. Along with the already identified two seed bits, this new information is utilized in solving linear equations; all the remaining three bits of the seed are recovered. Our attack can actually be terminated successfully at that point, although two more iterations are shown in the table. 

\begin{table}[!tb]
\centering
\caption{ScanSAT on dynamic scan locking for $p$=1. The seed 00001 is recovered upon two independent SAT attack runs.}
\label{tab:dyn_lock}
\resizebox{0.45\textwidth}{!}{
\begin{tabular}{@{}cccc@{}}
\toprule
Dynamic key & Correct value & Recovered key & Recovered seed \\ \midrule
key1 & 10000 & -00--- & 00---- \\ \midrule
key2 & 01000 & 010-0 & 00001 \\ \midrule
key3 & 00100 & 001-- & 00001 \\ \midrule
key4 & 10010 & 10010 & 00001 \\ \bottomrule
\end{tabular}
}
\end{table}

\section{ScanSAT on Scan Compression}

The modeling equations developed in the previous sections assume that the DfT structure has no scan compression. 
In this section, we elaborate on how ScanSAT attack can be extended to scan architectures with scan compression. The discussion herein applies to both static and dynamic scan obfuscation. For simplicity of discussion, and without loss of generality, we utilize fanout decompression and XOR compaction as example stimulus decompression and response compaction, respectively, to explain our attack. We note that the proposed attack can also be applied for any other stimulus decompression and response compaction technique. 

With multiple obfuscated scan chains, the same modeling technique in Fig.~\ref{fig:encrupt_scan_chain_xor} can be applied with no changes. The resulting equations, though, would be simpler than those for the single scan chain case, as the cascaded effect of each key-bit is limited by the depth of the chain where the key-bit is inserted. 

\begin{figure}[!t]
	\centering
	\includegraphics[width=0.45\textwidth]{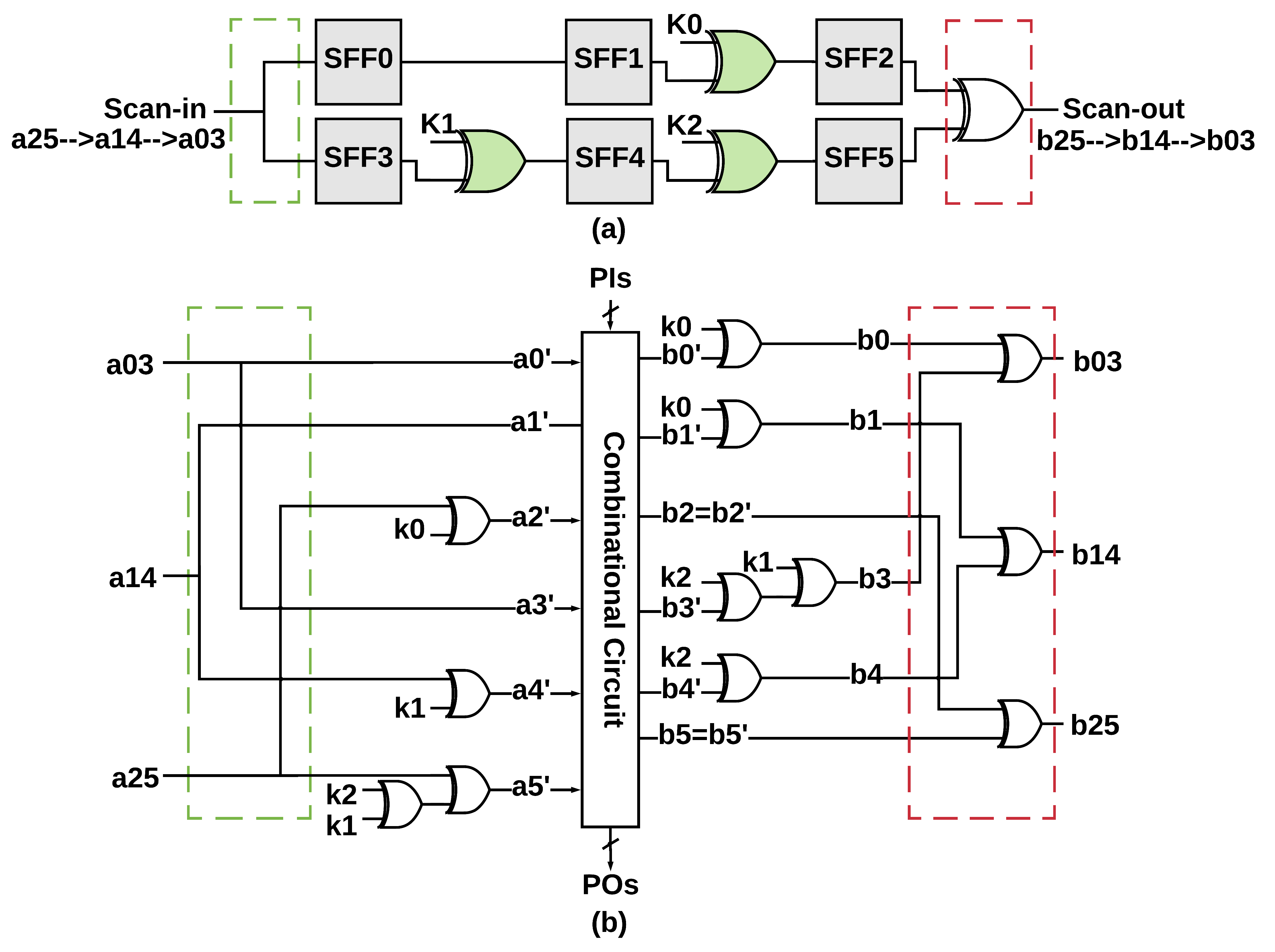}
	\caption{(a) Applying scan obfuscation on a compression-based architecture; compression ratio is 2 and three key-bits are used. (b) Modeling as a logic locking problem.}
	\label{fig:compression_model_example}
\end{figure}

With stimulus decompressor and response compactor around the obfuscated scan chains, there is one additional step for ScanSAT modeling; the decompressor and compactor structures need to be instantiated as many times as the number of scan slices, capturing the decompression into and compaction of individual slices.\footnote{In scan compression, the group of flip-flops that receive their stimulus in the same shift cycle is referred to as a {\em scan slice}. The number of scan slices is also referred to as the {\em scan depth}.} The modeled circuit then relates the compressed stimulus to delivered stimulus and captured response to observed (compacted) response, both through key-bits. 

The scan architecture in Fig.~\ref{fig:compression_model_example} (b) assumed on the circuit shown in Fig.~\ref{fig:encrupt_scan_chain_xor} results in the scan obfuscated architecture in Fig.~\ref{fig:compression_model_example} (a). In this example, three stimulus bits $a_{25}$, $a_{14}$, and $a_{03}$ are delivered into the six SFFs in three shift cycles; some of these stimulus bits are inverted as they pass through the key gates. Upon capture, the six captured response bits are compacted into three bits $b_{25}$, $b_{14}$, and $b_{03}$ to be observed through the Scan-out pin; some of these six bits are inverted prior to compaction as they pass through the key gates. 

Let us assume a combinational decompression that implements a function $D_{j \in IN(i)}(a_{j})$ for scan cell $i$ to compute $a_i$ from the compressed scan-in stimulus bits $a_j$. With scan obfuscation in place, the bits delivered into the SFFs in terms of the scan-in stimulus and the key-bits are as follows:

\begin{equation}
a_{i}^{'}= D_{j \in IN(i)}(a_{j}) \oplus L(k,i)
\label{eq:ai_comp}
\end{equation}

The one-to-one correspondence between the scan-in pattern bits and the delivered pattern bits was captured via Equation \ref{eq:ai}; this one-to-one correspondence no longer exists due to scan compression as illustrated by Equation \ref{eq:ai_comp}. The end-result is reduced controllability of $a^{'}$, which is expected due to stimulus decompression.

The delivered pattern $a^{'}$ applied on the combinational circuit produces the response captured in the SFFs $b^{'}$. These bits pass through key-bit driven key gates prior to compaction and being observed as three compacted bits $b_{03}$, $b_{14}$, and $b_{25}$ in Fig.~\ref{fig:compression_model_example} (a). Let us assume a combinational compactor that implements a function $C_{i \in IN(j)}(b_{i}^{'})$ to compute the compacted response bit $b_j$ from the captured response bits $b_i^{'}$. The following equations model the obfuscation and compaction operations:

\begin{equation}
b_{j}= C_{i \in IN(j)}(b_{i}^{'} \oplus R(k,i))
\label{eq:bi_comp}
\end{equation}

The one-to-one correspondence between the captured response bits and the observed response bits was captured via Equation \ref{eq:bi}; this one-to-one correspondence no longer exists due to response compaction as illustrated by Equation \ref{eq:bi_comp}. The end result is reduced observability of $b^{'}$, which is expected due to response compaction.

The ScanSAT modeled circuit for this example that captures the equations above is provided in Fig.~\ref{fig:compression_model_example} (b). This modeled circuit shows the fanout decompressor and the XOR compactor each instantiated three times in order to model the stimulus decompression into the three slices and the response compaction of the three slices. The final modeled circuit in Fig.~\ref{fig:compression_model_example} (b) is again a logic-locked circuit that has three key-bits, which the attacker can break by applying the SAT attack. This time, the input patterns $a$ are compressed scan-in patterns that the attacker applies from the Scan-in pin of a working chip. The output patterns $b$ are the compacted response patterns that the attacker collects from the Scan-out pin of the working chip. The reduced controllability due to stimulus compression and reduced observability due to response compaction may reflect into increased attack difficulty; the computation of the key values is now subject to these controllability and observability challenges. The more aggressive the compression ratio, the more difficult the attack may become.

\section{Experimental Results}
\subsection{Experimental Setup}
In this section, we present the experimental results for the ScanSAT attack on several circuits locked using static and dynamic scan obfuscation, with Encrypt Flip-Flop~\cite{karmakar2018encrypt}, \color{black} scan scrambling~\cite{hely2004scan}, \color{black} and DOS~\cite{wang2017secure} used as representative examples. We implemented Encrypt Flip-Flop\footnote{While the preliminary version of this work in~\cite{aspdac19} implemented Encrypt Flip-Flop as is (static scan locking and logic locking integrated together), in this paper, we implemented only the static scan locking part of it for consistency of comparisons. We do have experiments where static scan locking is coupled with logic locking in this paper as well.}, \color{black} scan scrambling, \color{black} DOS, and ScanSAT in a Perl framework on the largest four circuits, i.e., $s38584$, $s38417$, $s35932$, and $b19$, from ISCAS-89~\cite{iscas89} and ITC-99 benchmarks~\cite{itc99}. The details of the benchmark circuits are listed in Table \ref{tb:benchmark}. 
The benchmark circuits equipped with the full scan infrastructure are locked using 128-bit scan locking keys. 

For launching ScanSAT on the obfuscated scan chains with compression infrastructure, 
the SFFs in a design are configured into 16 scan chains and compression ratios (R) of 1, 2, 4, 8, and 16 are used. All the tests were performed on Ubuntu virtual machine utilizing one core of Intel $i7-3770$ CPU running at $3.40GHz$ with $10$ GB of RAM.
\subsection{ScanSAT Attack on Static Scan Obfuscation}

\begin{table}[!t]
\centering
\caption{Statistics of the benchmark circuits~\cite{iscas89,itc99}. The circuits are sorted based on the number of SFFs.}
\footnotesize
\label{tb:benchmark}
\begin{tabular}{@{}ccccc@{}}
\toprule
\multicolumn{5}{c}{ISCAS-89\textbackslash ITC-99 Benchmark}              \\ \midrule
Circuit & \#SFFs & \#inputs & \#outputs & \#gates \\ \midrule
s38584  & 1426   & 38       & 304       & 19253   \\ \midrule
s38417  & 1636   & 28       & 106       & 22179    \\ \midrule
s35932  & 1728   & 35       & 320       & 16065   \\ \midrule
b19     & 6642   & 24       & 30       & 231320\\\bottomrule
\end{tabular}
\end{table}

ScanSAT attack results on statically obfuscated scan architectures with \textbf{scan compression} are listed in Table~\ref{tb:compression_table}. {\bf ScanSAT is successful in 100\% of the cases and retrieves the correct key value for all the circuits}.
The vulnerability of scan chain obfuscation to the proposed ScanSAT is demonstrated by the fact that only a few DIPs are required to unlock the circuits even with a key size of 128. 
We attribute this extremely low number of DIPs to the ability of the proposed modeling to efficiently capture the data dependencies in the scan chain. As illustrated earlier in Fig.~\ref{fig:model_example}, each key-bit affects (i) the stimulus delivered to a scan cell to the right of it and (ii) the response collected from a scan cell to the left of it. Thus, the error introduced by any incorrect key-bit is expected to have a unique impact, with the exception of errors being masked during the capture operation. Easy distinguishability of keys results in a very effective key pruning via the SAT attack.

The results in Table~\ref{tb:compression_table} also confirm that with scan compression in place, more aggressive compression ratios generally reflect into increased attack times. The underlying reason, as mentioned earlier, is the reduced controllability and observability during the SAT attack. 

\begin{table}[!t]
\centering
\footnotesize
\caption{ScanSAT attack results on statically obfuscated scan chains with compression. All benchmarks are locked with a key size of 128 and 16 scan chains are constructed in each design. $R=1$ implies that there is no compression.}
\label{tb:compression_table}

\begin{tabular*}{0.45\textwidth}{@{}c @{\extracolsep{\fill}} cccc@{}}
\toprule
Circuit & \#SFFs & \begin{tabular}[c]{@{}c@{}}Compression\\ ratio (R)\end{tabular} & \#DIPs & \begin{tabular}[c]{@{}c@{}}Execution\\ time (s)\end{tabular} \\ \midrule

\multirow{5}{*}{s38584} & \multirow{5}{*}{1426} & 1 & 1 & 20 \\ \cmidrule(l){3-5} 
 &  & 2 & 1 & 16 \\ \cmidrule(l){3-5} 
 &  & 4 & 1 & 14 \\ \cmidrule(l){3-5} 
 &  & 8 & 1 & 33 \\ \cmidrule(l){3-5} 
 &  & 16 & 1 & 40 \\ \midrule
\multirow{5}{*}{s38417} & \multirow{5}{*}{1636} & 1 & 8 & 30 \\ \cmidrule(l){3-5} 
 &  & 2 & 7 & 40 \\ \cmidrule(l){3-5} 
 &  & 4 & 9 & 65 \\ \cmidrule(l){3-5} 
 &  & 8 & 5 & 74 \\ \cmidrule(l){3-5} 
 &  & 16 & 7 & 701 \\ \midrule
\multirow{5}{*}{s35932} & \multirow{5}{*}{1728} & 1 & 1 & 9 \\ \cmidrule(l){3-5} 
 &  & 2 & 1 & 12 \\ \cmidrule(l){3-5} 
 &  & 4 & 1 & 13 \\ \cmidrule(l){3-5} 
 &  & 8 & 1 & 13 \\ \cmidrule(l){3-5} 
 &  & 16 & 1 & 17 \\ \midrule
\multirow{5}{*}{b19} & \multirow{5}{*}{6642} & 1 &15 & 5771 \\ \cmidrule(l){3-5} 
 &  & 2 &19 & 26726 \\ \cmidrule(l){3-5} 
 &  & 4 &  15 &  28908 \\ \cmidrule(l){3-5} 
 &  & 8 &13 & 18156  \\ \cmidrule(l){3-5} 
 &  & 16 &  12 & 12496 \\ \bottomrule

\end{tabular*}
\end{table}

\subsection{ScanSAT on Scan Chain Scrambling}
\label{sec:exp_scramble}
In this section, we present the experimental results for the ScanSAT attack on circuits locked using scan chain scrambling in addition to static scan obfuscation. The correct scan paths are generated randomly, and then based on those generated paths, new routing logic is added. The correct key value is the one that ensures the specified scan paths. The benchmark circuits equipped with the full scan infrastructure are scrambled with a fixed key size of 64, and then the benchmarks are additionally locked using static obfuscation with 128-bit keys. ScanSAT attack results on the scrambled scan architectures are listed in Table~\ref{tb:scramble_result}. {\bf Again, the attack is successful in 100\% of the cases.} A scan obfuscation key of 0 means that the circuit is locked using the scrambling technique only. The proposed ScanSAT attack is capable of consistently breaking scan scrambling with or without static scan locking used in conjunction.

\begin{table}[!t]
\centering
\caption{\color{black}ScanSAT attack results on scan chain scrambling with static scan obfuscation.\color{black}}
\label{tb:scramble_result}
\resizebox{0.45\textwidth}{!}{%
\begin{tabular}{@{}ccccccc@{}}
\toprule
Circuit & \begin{tabular}[c]{@{}c@{}}\#Scan\\ chains\end{tabular} & \begin{tabular}[c]{@{}c@{}}Scrambling\\ key\end{tabular} & \begin{tabular}[c]{@{}c@{}}Obfuscation\\ key\end{tabular} & \begin{tabular}[c]{@{}c@{}}Total\\ key size\end{tabular} & \#DIPs & \begin{tabular}[c]{@{}c@{}}Execution\\ time (s)\end{tabular} \\ \midrule
\multirow{4}{*}{s38584} & \multirow{4}{*}{9} &  & 0 & 64 & 5 & 41 \\ \cmidrule(l){4-7} 
 &  &  & 128& 192 & 2 & 108 \\\cmidrule(r){1-2} \cmidrule(l){4-7} 
\multirow{4}{*}{s38417} & \multirow{4}{*}{17} & \multirow{6}{*}{64} & 0 & 64 & 2 & 27 \\ \cmidrule(l){4-7} 
 &  &  &128& 192 & 6 & 80 \\ \cmidrule(r){1-2} \cmidrule(l){4-7} 
\multirow{4}{*}{s35932} & \multirow{4}{*}{16} &  & 0 &64 & 2 & 50 \\ \cmidrule(l){4-7} 
 &  &  &128& 192 & 3 & 78 \\\cmidrule(r){1-2} \cmidrule(l){4-7} 
\multirow{4}{*}{b19} & \multirow{4}{*}{16} &  & 0 &64 & 8 & 30439   \\ \cmidrule(l){4-7} 
 & & &128& 192 & 13 & 17134 \\  \bottomrule
\end{tabular}
}

\end{table}
\color{black}
\subsection{ScanSAT on Dynamic Scan Obfuscation}
\label{sec:exp_dynamic}

\color{black} We implement the most challenging case of dynamic scan locking, i.e., for the case $p$=1; the dynamic keys are updated in every cycle. We then apply the proposed Scan SAT on this challenging defense. We report the number of iterations (i.e., the number of independent SAT attack runs), which is the same as the total number of DIPs generated by ScanSAT, as one DIP is generated in each iteration ($p$=1). \color{black}

\subsubsection{Detecting the Key Update Frequency}
The first set of experiments were conducted in order to prove that the key update frequency $p$ can be detected by repeatedly inserting the same scan-in pattern(s), performing capture operations, and observing the scan-out pattern(s). 
We applied our attack on all the benchmarks. \textbf{ScanSAT is successful in 100\% of the cases as it retrieves the correct key update frequency value across all compression ratios.}

\subsubsection{Retrieving the LFSR Seed}
\begin{figure}[!t]
	\centering
	\includegraphics[width=0.5\textwidth]{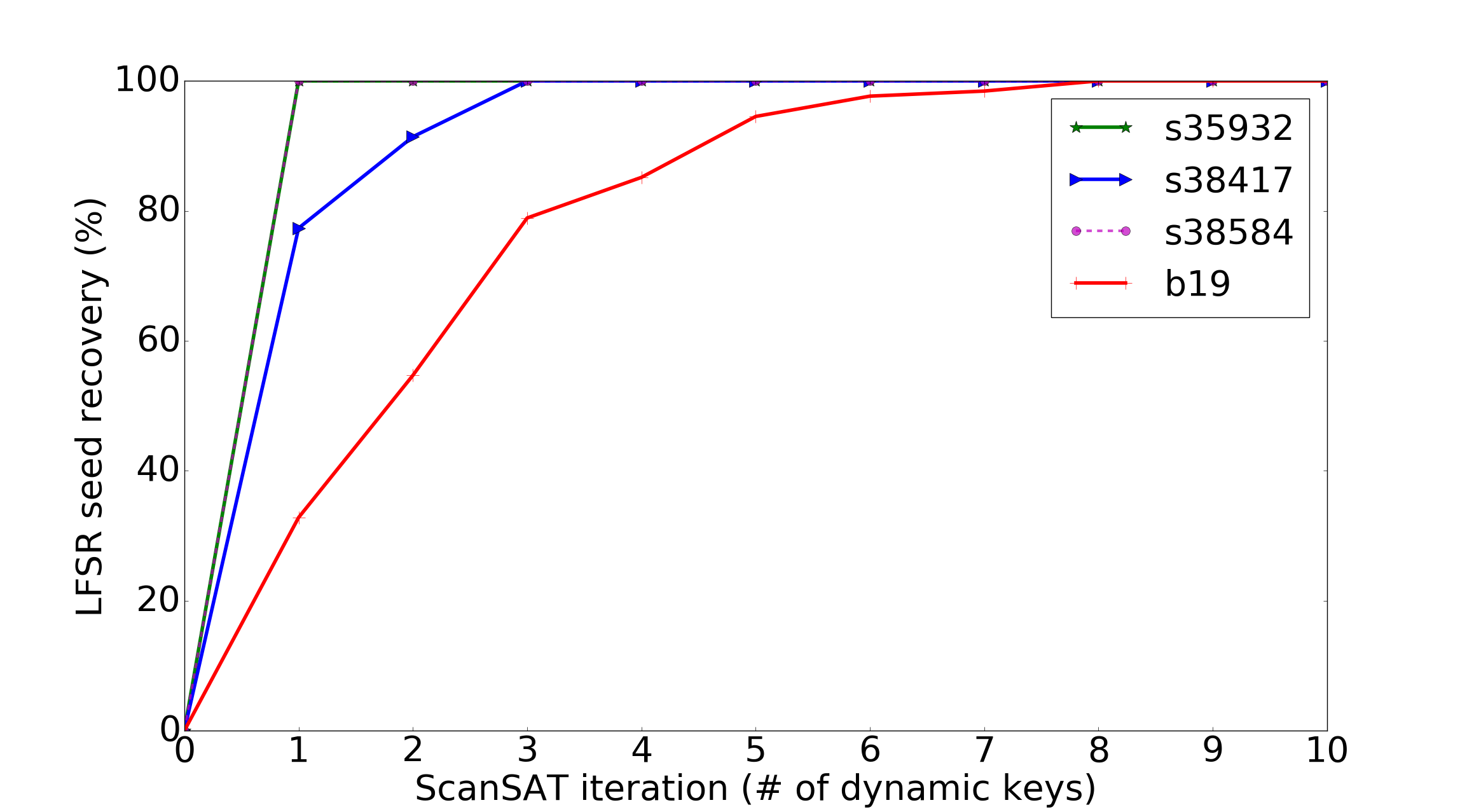}
	\caption{\color{black}ScanSAT on dynamic scan locking with no compression. Percentage of seed recovered in every ScanSAT iteration. In every iteration, a new dynamic key is active as $p$=1.}
	\label{fig:dynlock_graph}
\end{figure}

\color{black} A 128-bit LFSR was modeled as a combinational circuit in order to obtain the seed-to-key block in our attack.  Fig.~\ref{fig:dynlock_graph} shows the percentage of seed bits gradually recovered with each iteration of the proposed ScanSAT attack. We can observe that it takes only eight iterations for the proposed attack to break dynamic scan locking even for $p$=1 to recover the complete LFSR seed for the $b19$ benchmark circuit. For circuits $s35932$ and $s38584$, we recover the seed within only one iteration.

Table~\ref{tb:dyn_compression_table} presents the number of iterations and time needed in order to break dynamic scan obfuscation when scan compression is in place. Comparing the obtained results again with the results presented in Table~\ref{tb:compression_table}, it can be noted that there is no consistent trend in the execution time required to unlock the circuits that are statically vs dynamically obfuscated. The reason is that a single SAT attack run with many DIPs (attack on static scan locking) is not that different in terms of execution time than multiple independent SAT attack runs with few DIPs (attack on dynamic scan locking) where the CNF is carried over from one run to the next. Moreover, the number of iterations required to unlock the $s38584$ and the $s35932$ circuits for different compression ratios is always one, same as in the static scan locking case, as expected.
{\bf In all the cases, ScanSAT uses only a few iterations to break dynamic obfuscation.} In a couple of cases, a few seed bits remain unsolved due to attack time-out; these are $s38417$ with a compression ratio of 4 (2 bits remain unsolved) and $b19$ with a compression ratio of 2 (9 bits remain unsolved). We assume these cases to be broken as well, since the remaining bits can easily be identified via brute-force. 

\begin{table}[!t]
\centering
\footnotesize
\caption{ScanSAT attack results on dynamically obfuscated scan chains with compression and p = 1. All benchmarks are locked with a key size of 128 and 16 scan chains are constructed in each design. $R=1$ implies that there is no compression.}
\label{tb:dyn_compression_table}

\begin{tabular*}{0.48\textwidth}{@{}c @{\extracolsep{\fill}} cccc@{}}
\toprule
Circuit & \begin{tabular}[c]{@{}c@{}}Compression\\ ratio (R)\end{tabular} & \begin{tabular}[c]{@{}c@{}}\#Dynamic\\ key\end{tabular} & \begin{tabular}[c]{@{}c@{}}Seed\\ recovered\end{tabular} & \begin{tabular}[c]{@{}c@{}}Execution\\ time (s)\end{tabular} \\ \midrule

\multirow{5}{*}{s38584}                                       & 1                                                                 & 1 & 128       & 72                                                  \\ \cmidrule(l){2-5} 
                                                              & 2                                                                 & 1 & 128      & 49                                 
\\\cmidrule(l){2-5} 
                                                              & 4                                                                 & 1  & 128     & 73                         
\\\cmidrule(l){2-5} 
                                                              & 8                                                                 & 1  & 128     & 89                                                   
\\ \cmidrule(l){2-5} 
                                                              & 16                                                                & 1  & 128    & 125                                                   
\\ \midrule
\multirow{5}{*}{s38417}                                      & 1                                                                 & 3 & 128       & 59                                                   \\ \cmidrule(l){2-5} 
                                                            & 2                                                                 & 7 & 128       & 231                                                   
                                                            \\ \cmidrule(l){2-5} 
                                                           & 4                                                                 & 7 & 126       & 292                                                  
                                                           \\ \cmidrule(l){2-5} 
                                                          & 8                                                                 & 5 & 128       & 175                                                 
                                                          \\ \cmidrule(l){2-5} 
                                                          & 16                                                                & 7 & 128       & 836                                                 
                                                          \\ \midrule                        
\multirow{5}{*}{s35932}                                        & 1                                                                 & 1 & 128      & 17                                                  \\ \cmidrule(l){2-5} 
                                                                & 2                                                                 & 1 & 128      & 17                                                   \\ \cmidrule(l){2-5} 
                                                                & 4                                                                 & 1 & 128      & 18                                                   \\ \cmidrule(l){2-5} 
                                                                & 8                                                                 & 1 & 128      & 48                                                  \\ \cmidrule(l){2-5} 
                                                                & 16                                                                & 1 & 128      & 54                                                   \\ \midrule                         
\multirow{5}{*}{b19}                                        & 1                                                                    & 8 & 128      & 828                                                  \\ \cmidrule(l){2-5} 
                                                           & 2                                                                 & 11 & 119      & 2296                                                   
                                                           \\ \cmidrule(l){2-5} 
                                                            & 4                                                                 & 11 & 128      & 2447                                              
                                                            \\ \cmidrule(l){2-5} 
                                                           & 8                                                                 & 11 & 128      & 4660                                                   
                                                           \\ \cmidrule(l){2-5} 
                                                           & 16                                                                & 8 & 128      & 10065                                                   
                                                           \\ \bottomrule                        
\end{tabular*}
\end{table}

\subsection{Scan Chain Obfuscation + Random Logic Locking}
\label{sec:RLL}
We next investigate the difficulty of breaking scan locking integrated with a combinational logic locking defense, such as Random Logic Locking (RLL), wherein XOR/XNOR key gates are inserted at random locations in the combinational circuit~\cite{epic}. The circuits with a compression ratio of 16 are locked using both static and dynamic scan obfuscation with a key size of $128$. An additional RLL layer is integrated with additional $128$ key-bits, locking the circuits with a total of $256$ key-bits. \color{black} \textbf{The original SAT attack, when launched on circuits that are scan-obfuscated and logic-locked, consistently fails in obtaining the logic locking key; the DIPs that should be applied to the oracle could not be applied intact due to the scan obfuscation in place, which the SAT attack failed to account for in generating its DIPs}. \color{black} ScanSAT is then launched on the scan-obfuscated and logic-locked circuits; the results are listed in Table~\ref{tb:with_RLL}. {\bf The attack is 100\% successful on all cases again for both static and dynamic scan obfuscation.} 

Comparing the obtained results with the results presented in Tables~\ref{tb:compression_table} and~\ref{tb:dyn_compression_table}, it can be noted that the execution time required to unlock the circuits becomes higher when a second RLL layer is in place. For the static scan-obfuscated $s35932$ circuit, for example, the attack took $1.4\times$ longer time to terminate compared to when no RLL was integrated; four DIPs are now utilized by the attack, whereas previously only one DIP was employed. This can be expected as the key size is now doubled. We conclude that integration of scan obfuscation as another layer of defense over a vulnerable logic locking technique is still vulnerable to ScanSAT for both static and dynamic scan obfuscation. {\bf This confirms our claim that SAT attack resilient logic locking solutions are still needed to protect design IP, as one cannot rely on scan obfuscation to protect logic locking.}

\begin{table}[!t]
\centering
\caption{ScanSAT attack results on statically/dynamically obfuscated (128 bits) scan chains with RLL (128 bits). Compression ratio is 16. Total key size is 256.}
\footnotesize
\label{tb:with_RLL}
\begin{tabular}{@{}ccccc@{}}
\toprule
\multirow{2}{*}{Circuit} & \multicolumn{2}{c}{Static obfuscation} & \multicolumn{2}{c}{Dynamic obfuscation} \\ \cmidrule(l){2-5} 
 & \#DIPs & Execution time (s) & \#Iter & Execution time (s) \\ \midrule
s38584 & 20 & 94 & 13& 554\\ \midrule
s38417 & 49 & 1927& 9 & 2567\\ \midrule
s35932 & 4 & 23& 3& 85\\ \midrule
b19 & 57 & 70972 &  9 & 29969\\ \bottomrule
\end{tabular}
\end{table}

\section{Discussion}
A recent survey paper~\cite{scan_locking_survey} lists various scan obfuscation/masking defenses. In this section, we categorize these defenses into three classes as shown in Table~\ref{tab:scansat_table}: (i) RE-vulnerable: those that are vulnerable to any basic reverse engineering (RE) attack; e.g., the position of added gates identified via RE can be used to compromise the defense; ScanSAT does not need to be applied. These are defenses that are more applicable in threat models where reverse engineering is considered impractical.
(ii) ScanSAT-vulnerable: those that are vulnerable to our attack ScanSAT.
(iii) ScanSAT-resilient: those that are resilient to our attack ScanSAT but at the expense of other implications such as hindered debug, etc. Details about these defenses can be found in~\cite{scan_locking_survey}. 

\subsection{ScanSAT versus Scan Attacks}  
ScanSAT and other oracle-guided attacks assume access to a reverse-engineered netlist (IP); such a netlist is not included in the threat model of the scan attacks. As a result, countermeasures for scan attacks are no longer secure in the more generous threat model.  
Certain scan attack countermeasures rely on scan chain  authentication~\cite{lee2007securing,cui2017static,xor_secure_scan_2012,paul2007vim}. 
These countermeasures can be circumvented by having access to a reverse-engineered netlist, as the on-chip logic that implements secure scan can also be reverse-engineered to bypass authentication. 

Also, most side-channel attacks rely on operating in normal/user mode for a few cycles and then switching to the test mode; the data loaded into the round registers in the user mode can leak through the SFFs during the test mode~\cite{yang2004scan}.
The Mode-Reset Countermeasure (MRC) resets all flip-flops upon transition from one mode to the other and thwarts traditional scan attacks~\cite{hely2005test}; the countermeasure is deployed in many Intel chips. 
MRC provides no protection against ScanSAT, as {\bf ScanSAT operates only in the test mode}. 

\subsection{Dynamicity of keys}

In this paper, we show that ScanSAT can break dynamic scan obfuscation for $p$=1. In order to break the improved EFF scheme in~\cite{eff_new}, we can tweak our modeling to account for the dynamicity of the key across shift cycles; then our attack would be able to retrieve the seed of the LFSR that generates these keys. Our future research will focus on this more challenging problem. We also note that significant practicality problems emerge as the key dynamicity becomes more aggressive; as explained in~\cite{wang2017secure}, even small values for p may not be that practical as test patterns are often post-processed (addition/removal/reordering of test patterns) while the dynamic keys are produced in a fixed sequence by the LFSR. The original dynamic scan obfuscation scheme in~\cite{wang2017secure} suggests the use of a large enough p that will allow for the re-use of the same key for the reordered test patterns within a window of p patterns.
\color{black}

\subsection{Limitations of ScanSAT} 
ScanSAT may fail against Built-in Self Test (BIST)~\cite{mccluskey1985built} due to the extremely limited observability; this expectation can be verified by extrapolating from the data in Tables~\ref{tb:compression_table} and~\ref{tb:dyn_compression_table} for the very large compression ratio of BIST. However, BIST also inhibits debugging and in-field testing the same way it inhibits our attack.

\color{black} Further, ScanSAT can be thwarted by encrypting/decrypting the stimulus and the response by using provably secure techniques~\cite{da2018preventing}; but this requires using dedicated, and thus, costly on-chip cipher blocks. \color{black} Another approach would be to utilize PUFs as in~\cite{da2018new} at the expense of reliability concerns associated with PUFs.\color{black}

Another way to thwart ScanSAT may be to decouple the LFSR update from the key update in dynamic scan obfuscation. The proposed ScanSAT can model such an operation, as the structural details of the LFSR (its polynomial) can be utilized to compute the relationship between the seed and the produced keys. The attack can still be applied as is and expected to be effective; from the identified linear equations, the seed can be computed by accounting for the LFSR states that are ``skipped." Keying the rate ratio between the LFSR to key update, however, would significantly hamper the attack; the exact relationship between the seed and the dynamic keys would be hidden from the attacker. Future research is necessary to enhance the proposed ScanSAT to model the seed to key relationship and circumvent such a defense.

\subsection{Comparison Against NSAA~\cite{massad17}}
\color{black}
An orthogonal line of research is an attack that assumes no scan access~\cite{massad17}, herein referred to as NSAA (no scan access attack). 
NSAA exercises only the primary inputs (and not the SFFs) and observes only the primary outputs (and not the SFFs) of the working chip.
In~\cite{massad17}, NSAA is reported to be successful for 80\% of the time for a key-size of 32 bits. Often, the attack can correctly retrieve only a subset of key-bits. In contrast, ScanSAT works 100\% of the time, even on large circuits such as $s38584$ and $b19$ (even with a key-size of 128).
For $s38584$, the  only  benchmark  common to our and their work,  the NSAA tool reportedly crashed~\cite{massad17}.

As acknowledged in~\cite{massad17}, NSAA is effective only if the ratio of the primary IOs to the SFFs is reasonably large. Unfortunately, this ratio is expected to be small for the realistic circuits as the chip interface (primary IOs) cannot grow at the same rate as the design complexity (SFFs). 
\begin{table}[!t]
\centering
\caption{\color{black} Scan obfuscation/masking defenses}
\label{tab:scansat_table}
\resizebox{0.48\textwidth}{!}{%
\begin{tabular}{@{}ll@{}}
\toprule
{\bf Defense} & {\bf Security} \\ \midrule
VIm-Scan~\cite{paul2007vim} & RE-vulnerable \\\midrule
SSTKR~\cite{sstkr} & RE-vulnerable \\\midrule
\begin{tabular}[c]{@{}l@{}}Scan chain authentication~\cite{lee2007securing,cui2017static},\\State dependent scan flip-flop based scan (SDSFF)~\cite{xor_secure_scan_2012}\end{tabular} & RE-vulnerable \\\midrule
Flipped scan~\cite{sengar2007secured}, XOR scan~\cite{xor_scan}, double feedback XOR scan~\cite{double_feedback} & RE-vulnerable \\\midrule
Test key integrated scan~\cite{lee2006low} & RE-vulnerable\\\midrule
DOS~\cite{wang2017secure} & ScanSAT-vulnerable \\\midrule
EFF~\cite{karmakar2018encrypt}, improved EFF~\cite{eff_new} & ScanSAT-vulnerable \\\midrule
Scan scrambling~\cite{hely2004scan} & ScanSAT-vulnerable \\\midrule
Mode-reset countermeasure~\cite{hely2005test} & ScanSAT-vulnerable\\\midrule
\begin{tabular}[c]{@{}l@{}}BIST~\cite{mccluskey1985built} , scan on-chip comparison~\cite{compaction1}/compaction~\cite{compaction2}, \\ Secure test wrapper (STW)~\cite{chiu2010secure}\end{tabular} & ScanSAT-resilient \\\midrule
Scan interface encryption~\cite{da2018preventing}& ScanSAT-resilient \\\midrule
Bias PUF based scan~\cite{da2018new} & ScanSAT-resilient \\\midrule
\end{tabular}%
}
\end{table}
\color{black}
\section{Conclusion}
Obfuscation of scan chains aims to protect against the untrusted testers; static and dynamic scan locking techniques obfuscate the scan operations, hiding the relationship between the scan-in and the delivered stimuli and the relationship between the captured and the scan-out responses. Static scan locking utilizes a single key while dynamic scan locking keeps updating its key, resulting in a sequence of keys that obfuscate scan operations.

In this paper, we propose the ScanSAT attack on obfuscated scan chains, extracting the secret key (sequence) and unlocking the circuit/scan chain. The attack is evaluated by analyzing the security of three representative scan obfuscation techniques--two static and one dynamic--over different scan chain architectures. ScanSAT models the obfuscated scan chains as a logic-locked combinational circuit, paving the way for the application of the powerful SAT attack to reveal the key (sequence), unlocking the scan chains, and thus, restoring access to the oracle. We show that ScanSAT can break naive scan locking techniques even for large key sizes and when scan compression is in place.

We also show that ScanSAT is capable of breaking compound defenses comprising scan locking and logic locking, as long as the logic locking defense is vulnerable to SAT attack. We therefore nullify the presumption that scan locking may protect logic locking by hindering the full scan access that SAT or other logic locking attacks need; we show that a logic locking technique that is resilient to SAT attack is still needed to protect the design IP.
\section*{Acknowledgment}
This publication is based upon work supported by the Khalifa University of Science and Technology under Award No$.$ $[$RC2-2018-020$]$, New York University Abu Dhabi Center for Cybersecurity and Intel Corporation.

\ifCLASSOPTIONcaptionsoff
  \newpage
\fi
\linespread{0.88}
{\small \bibliographystyle{IEEEtran}
\bibliography{Bibfile} }
\begin{IEEEbiography}
[{\includegraphics[width=1in,height=1.25in,clip,keepaspectratio]{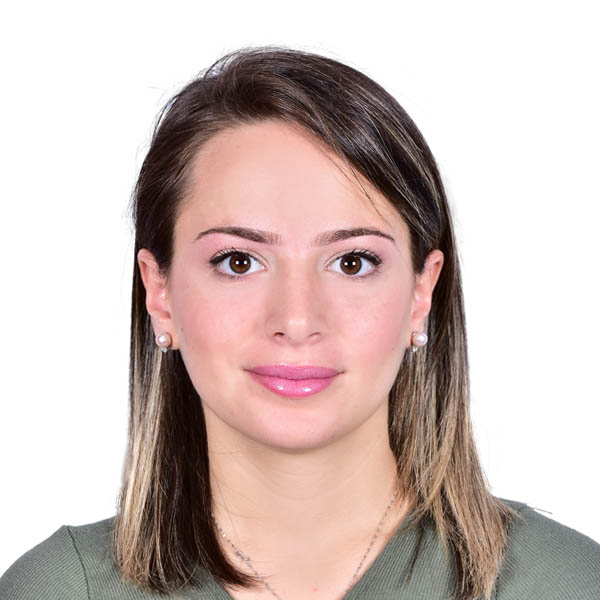}}]{Lilas Alrahis}
received the B.S. degree in Electrical and Electronics Engineering in 2014 and the M.S. degree in Electrical and Computer Engineering in 2016, both from Khalifa University, Abu Dhabi, UAE, where she is currently pursuing the Ph.D. degree in Engineering. Her current research interests include Hardware Security and Design for Trust.
\end{IEEEbiography}
\vskip -2\baselineskip  plus -1fil
\begin{IEEEbiography}
[{\includegraphics[width=1in,height=1.25in,clip,keepaspectratio]{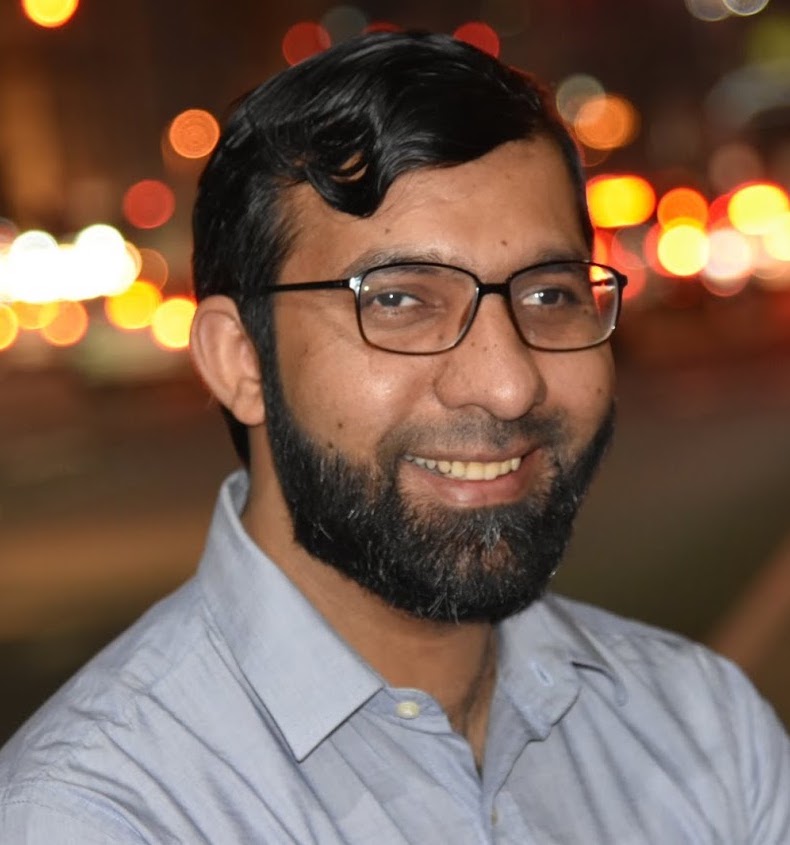}}]{Muhammad Yasin}
is a postdoctoral researcher in the Department of Electrical and Computer Engineering at Texas A\&M University. He obtained a Ph.D. degree in the Electrical and Computer Engineering Department at New York University in 2018; an MS in Microsystems Engineering from Masdar Institute of Science and Technology, UAE in 2013; and a BS in Electrical Engineering from University of Engineering and Technology (UET) Lahore, Pakistan in 2007. His research interests include Hardware Security, Design for Trust, and Logic Locking. He won the US semi-finals of the TTTC's E.J. McCluskey Best Doctoral Thesis Award in 2018.
\end{IEEEbiography}
\vskip -2\baselineskip plus -1fil
\begin{IEEEbiography}
[{\includegraphics[width=1in,height=1.25in,clip,keepaspectratio]{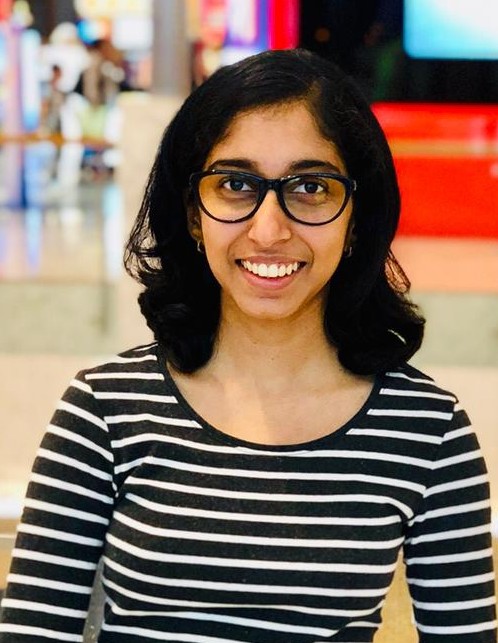}}]{Nimisha Limaye}
received B.E. in Electronics and Telecommunications Engineering from University of Mumbai, India in 2015 and M.S. in Computer Engineering from New York University, USA in 2017. She is a Ph.D. candidate at the Department of Electrical and Computer Engineering at New York University, USA and is also a Global Ph.D. Fellow with New York University Abu Dhabi, UAE. Her research interests include security of reversible computing and hardware security with particular focus on security of sequential circuits. She is a member of IEEE.
\end{IEEEbiography}
\vskip -2\baselineskip plus -1fil
\begin{IEEEbiography}
[{\includegraphics[width=1in,height=1.25in,clip,keepaspectratio]{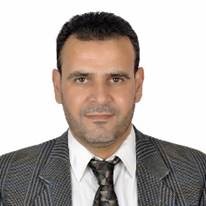}}]{Hani Saleh}
is an associate professor of electronic engineering at Khalifa University since 2012. He is a co-founder/active researcher in the KSRC (Khalifa University Research Center) and the System on Chip Research Center (SOCC). Hani has a total of 19 years of industrial experience in ASIC chip design. 

Prior to academia Hani worked for many leading chip design companies including Apple, Intel, AMD, Qualcomm, Synopsys, Fujitsu and Motorola. 

Hani received a Bachelor of Science degree in Electrical Engineering from the University of Jordan, a Master of Science degree in Electrical Engineering from the University of Texas at San Antonio, and a Ph.D. degree in Electrical and Computer Engineering from the University of Texas at Austin. Hani research interest includes IoT Devices design, deep learning, DSP algorithms design, computer. Hani has 16 issued US patents, 6 pending patent application, and over 100 articles published in peer reviewed conferences and Journals.

\end{IEEEbiography}
\vskip -2\baselineskip plus -1fil
\begin{IEEEbiography}
[{\includegraphics[width=1in,height=1.25in,clip,keepaspectratio]{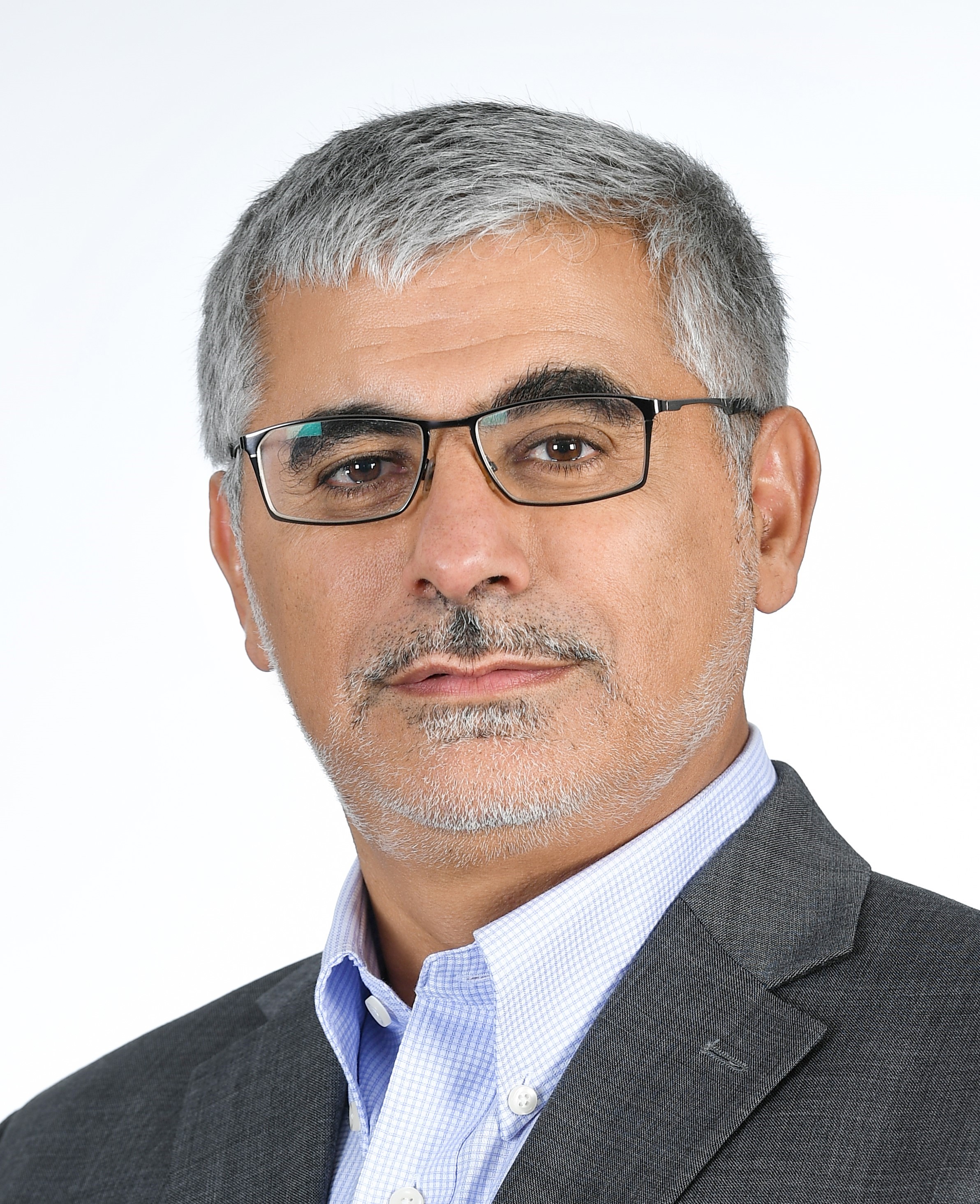}}]{Baker Mohammad}
earned his PhD from University of Texas at Austin, his M.S. degree from Arizona State University, Tempe, and BS degree from the University of New Mexico, Albuquerque, all in ECE.  Baker is the director of System on Chip Center and associate professor of ECE at Khalifa University.  Prior to joining Khalifa University Baker was a Senior staff Engineer/Manager at Qualcomm, Austin, USA for 6-years, where he was engaged in designing high performance and low power DSP processor used for communication and multi-media application. Before joining Qualcomm he worked for 10 years at Intel Corporation on a wide range of micro-processors design from high performance, server chips $>100$Watt (IA-64), to mobile embedded processor low power sub 1 watt (xscale).  His research interest includes VLSI, power efficient computing, Design with emerging technology such as Memristor, STTRAM, In-Memory-Computing, Efficient hardware accelerator for search engine, image processing, and Artificial Intelligent hardware.
\end{IEEEbiography}
\vskip -1\baselineskip plus -1fil
\begin{IEEEbiography}
[{\includegraphics[width=1in,height=1.25in,clip,keepaspectratio]{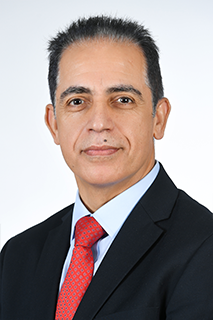}}]{Mahmoud Al-Qutayri}
received the B.Eng. degree from Concordia University, Canada, 1984, the M.Sc. degree from the University of Manchester, U.K., 1987, and the Ph.D. degree from the University of Bath, U.K., 1992 all in electrical and electronic engineering. He is currently a Full Professor with the Department of Electrical and Computer Engineering and the Associate Dean for Graduate Studies, College of Engineering at Khalifa University, UAE. Prior to joining Khalifa University, he worked at De Montfort University, UK and University of Bath, UK. He has authored/co-authored numerous technical papers in peer-reviewed journals and international conferences.  He also co-authored a book entitled Digital Phase Lock Loops: Architectures and Applications and edited a book entitled Smart Home Systems. This is in addition to a number of book chapters and patents.  Al-Qutayri current research interests include wireless sensor networks, embedded systems design, in-memory computing, mixed-signal integrated circuits design and test, and hardware security. 
\end{IEEEbiography}
\vskip -1\baselineskip plus -1fil
\begin{IEEEbiography}
[{\includegraphics[width=1in,height=1.25in,clip,keepaspectratio ]{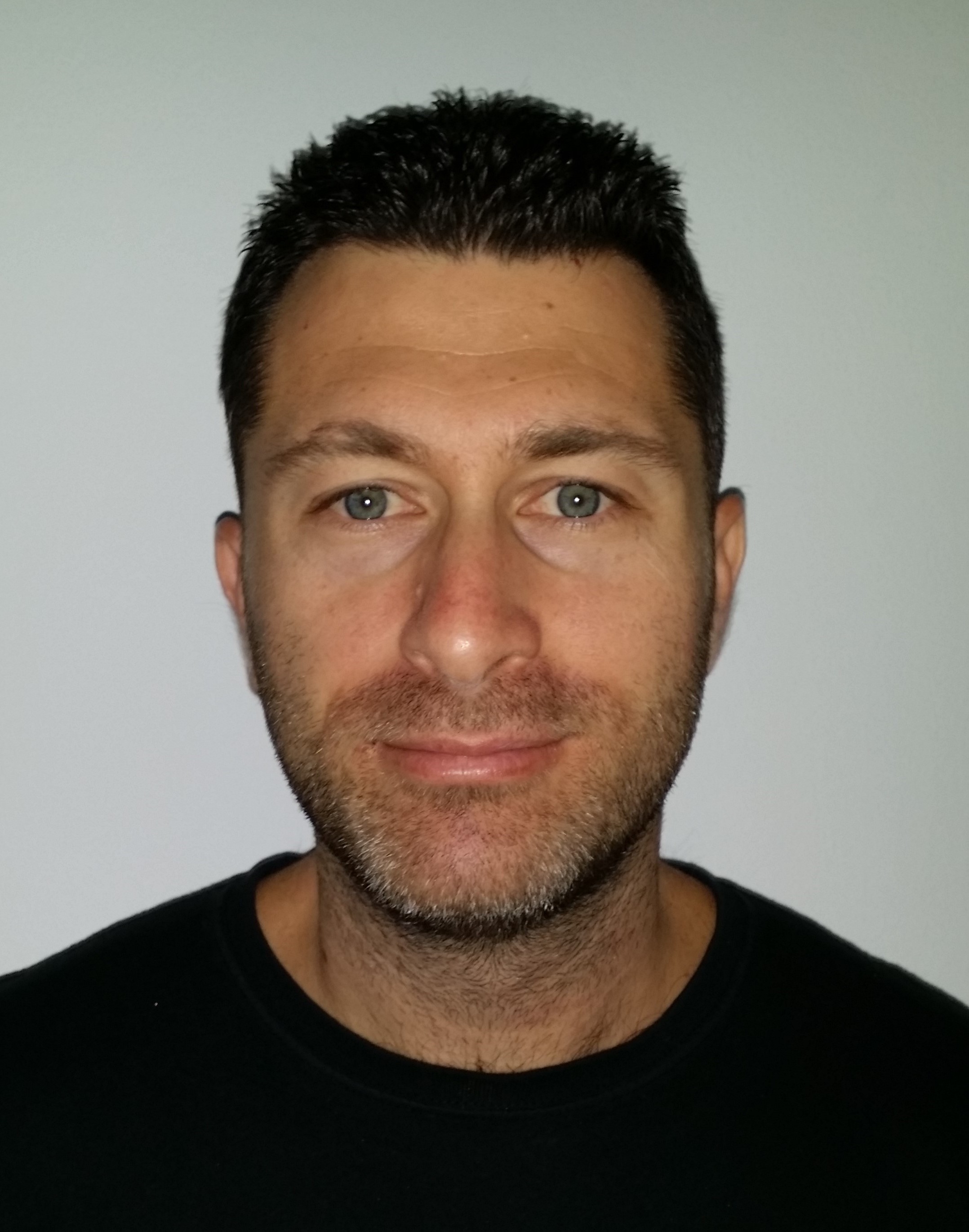}}]{Ozgur Sinanoglu}
is a professor of electrical and computer engineering at New York University Abu Dhabi. He obtained his PhD in Computer Science and Engineering from University of California San Diego in 2004. He has industry experience at TI, IBM and Qualcomm, and has been with NYU Abu Dhabi since 2010. During his PhD, he won the IBM PhD fellowship award twice. He is also the recipient of the best paper awards at IEEE VLSI Test Symposium 2011 and ACM Conference on Computer and Communication Security 2013. Prof. Sinanoglu’s research interests include design-for-test, design-for-security and design-for-trust for VLSI circuits, where he has around 200 conference and journal papers, and 20 issued and pending US Patents. His recent research in hardware security and trust is being funded by US National Science Foundation, US Department of Defense, Semiconductor Research Corporation, Intel Corp and Mubadala Technology.
\end{IEEEbiography}

\end{document}